\documentclass[a4paper,11pt]{article}
\pdfoutput=1 % if your are submitting a pdflatex (i.e. if you have
\usepackage{jinstpub} % for details on the use of the package, please
\usepackage{subfigure}
\usepackage{diagbox}
\usepackage{lineno}
\title{The Continuous Readout Stream of the MicroBooNE Liquid Argon Time Projection Chamber for Detection of Supernova Burst Neutrinos }
\collaboration{MicroBooNE Collaboration}

\author[ii]{P.~Abratenko}
\author[o]{M.~Alrashed}
\author[n]{R.~An}
\author[d]{J.~Anthony}
\author[hh]{J.~Asaadi}
\author[s]{A.~Ashkenazi}
\author[ll]{S.~Balasubramanian}
\author[k]{B.~Baller}
\author[t]{C.~Barnes}
\author[x]{G.~Barr}
% ...

% ...
\affiliation[d]{University of Cambridge, Cambridge CB3 0HE, United Kingdom}
% ...
\affiliation[k]{Fermi National Accelerator Laboratory (FNAL), Batavia, IL 60510, USA}
% ...
\affiliation[n]{Illinois Institute of Technology (IIT), Chicago, IL 60616, USA}
\affiliation[o]{Kansas State University (KSU), Manhattan, KS, 66506, USA}
% ...
\affiliation[s]{Massachusetts Institute of Technology (MIT), Cambridge, MA, 02139, USA}
\affiliation[t]{University of Michigan, Ann Arbor, MI, 48109, USA}
% ...
\affiliation[x]{University of Oxford, Oxford OX1 3RH, United Kingdom}
% ...
\affiliation[hh]{University of Texas, Arlington, TX, 76019, USA}
\affiliation[ii]{Tufts University, Medford, MA, 02155, USA}
% ...
\affiliation[ll]{Wright Laboratory, Department of Physics, Yale University, New Haven, CT, 06520, USA}

  \emailAdd{microboone\_info@fnal.gov}
\date{}

\abstract{The MicroBooNE continuous readout stream is a parallel readout of the MicroBooNE liquid argon time projection chamber (LArTPC) which enables detection of non-beam events such as those from a supernova neutrino burst. 
The low energies of the supernova neutrinos and the intense cosmic-ray background flux due to the near-surface detector location makes triggering on these events very challenging. 
Instead, MicroBooNE relies on a delayed trigger generated by SNEWS (the Supernova Early Warning System) for detecting supernova neutrinos.
The continuous readout of the LArTPC generates large data volumes, and requires the use of real-time compression algorithms (zero suppression and Huffman compression) implemented in an FPGA (field-programmable gate array) in the readout electronics.
We present the results of the optimization of the data reduction algorithms, and their operational performance. 
To demonstrate the capability of the continuous stream to detect low-energy electrons, a sample of Michel electrons from stopping cosmic-ray muons is reconstructed and compared to a similar sample from the lossless triggered readout stream.}

\begin{document}
\graphicspath{ {./figures/} }

\maketitle
\flushbottom

\section{Introduction}
Liquid argon time projection chamber (LArTPC) detectors capture particle interactions with exquisite spatial and calorimetric resolution producing large data volumes. For a typical ADC sampling rate of $2~\rm{MHz}$ and an ADC resolution of 12~bits, each channel generates $3~\rm{MB/s}$, without compression. In order to achieve good spatial resolution, most detectors use a wire pitch of a few millimeters spanning several meters, leading to several thousands of readout channels. Consequently, front-end electronics data rates can reach several $\rm{GB/s}$. 
For the acquisition of events from a neutrino beam, these data rates are manageable since the readout of the detector is driven by the accelerator beam spills, which occur with a known maximum frequency of $\mathcal{O}(10)~\rm{Hz}$, determined by the accelerator repetition rate, and only requires an acquisition window of milliseconds, determined by the electron drift time across the TPC.
For non-beam events such as supernova neutrino interactions or a search for nucleon decays that cannot be anticipated, it is a formidable task to process all of the continuous data, either for generating triggers based only on TPC information, or for data acquisition. 
Furthermore, with the upcoming multi-kiloton-scale detectors like the DUNE far detector modules~\cite{Abi:2018dnh} read out by hundreds of thousands of electronic channels, the challenge becomes more acute as data rates at the front-end will be of the order of~$\rm{TB/s}$. In addition, the capability of processing continuous data would enable the use of low-energy signals from radiological sources as calibration signals.

Within this scenario, the MicroBooNE detector, as the first LArTPC in operation with continuous readout, has a great opportunity to spearhead the development of the required technology and inform future experiments. The implementation of the continuous readout of the MicroBooNE LArTPC is described in section~\ref{sec:TheMicroBooNEContinuousReadoutStream}. The data reduction algorithms required to achieve the continuous readout within the MicroBooNE DAQ constraints are presented in section~\ref{sec:DataReductionAlgorithms}. The configuration of the main algorithm for zero suppression (ZS) is discussed in section~\ref{sec:ZeroSuppressionThresholds}. The data compression results achieved are presented in section~\ref{sec:DataRatesAndCompression}. The performance of the MicroBooNE continuous readout stream in detecting and reconstructing Michel electrons from stopping cosmic-ray muons, which are similar in energy to the ones expected from electron-neutrino interactions emitted by a core-collapse supernova, is discussed in section~\ref{sec:Analysis}. The potential of the MicroBooNE continuous readout stream as a development platform for other experiments such as DUNE is briefly discussed in section~\ref{sec:DevelopmentPlatform}.

\section{The MicroBooNE Continuous Readout Stream}
\label{sec:TheMicroBooNEContinuousReadoutStream}

The primary goal of the continuous readout stream is to enable the detection of signals from the burst of supernova neutrinos by the MicroBooNE detector, should a nearby core-collapse supernova happen during the lifetime of the experiment. For this reason, it is also known as the supernova stream (SN stream, used for the rest of the article). The SN stream is implemented in parallel to a triggered readout stream used for beam-related physics. 
Detection of supernova neutrinos in a LArTPC is especially interesting due to the higher sensitivity to the electron-neutrino flux through the $\nu_{e}+{\rm ^{40}Ar}\rightarrow e^{-}+{\rm ^{40}K^{*}}$ channel~\cite{Bueno:2003ei}.
The expected number of interactions in the MicroBooNE active volume is $\mathcal{O}(10)$ for a supernova burst at $10~\rm{kpc}$ (based on the prediction for DUNE~\cite{Abi:2018dnh}), spread over $\approx 10~\rm{s}$ and with energies in the range $\approx 5 - 50~\rm{MeV}$. 
Due to the large cosmic-ray rate ($\approx 5.5~\rm{kHz}$~\cite{Acciarri:2017rnj}) resulting from the close-to-surface location of the detector, MicroBooNE cannot rely on self-triggering on these events. Instead, the TPC data is continuously saved to disk on the DAQ servers and an alert issued by the Supernova Early Network System (SNEWS)~\cite{Antonioli:2004zb} can be used as a delayed trigger. 
The most recent data is kept for more than 48 hours, allowing the collaboration to react to the alert.
We defer the discussion about the reconstruction and selection of the supernova neutrino interactions in the MicroBooNE detector to future work.

\begin{figure}[htbp]
\centering
\includegraphics[width=0.98\textwidth]{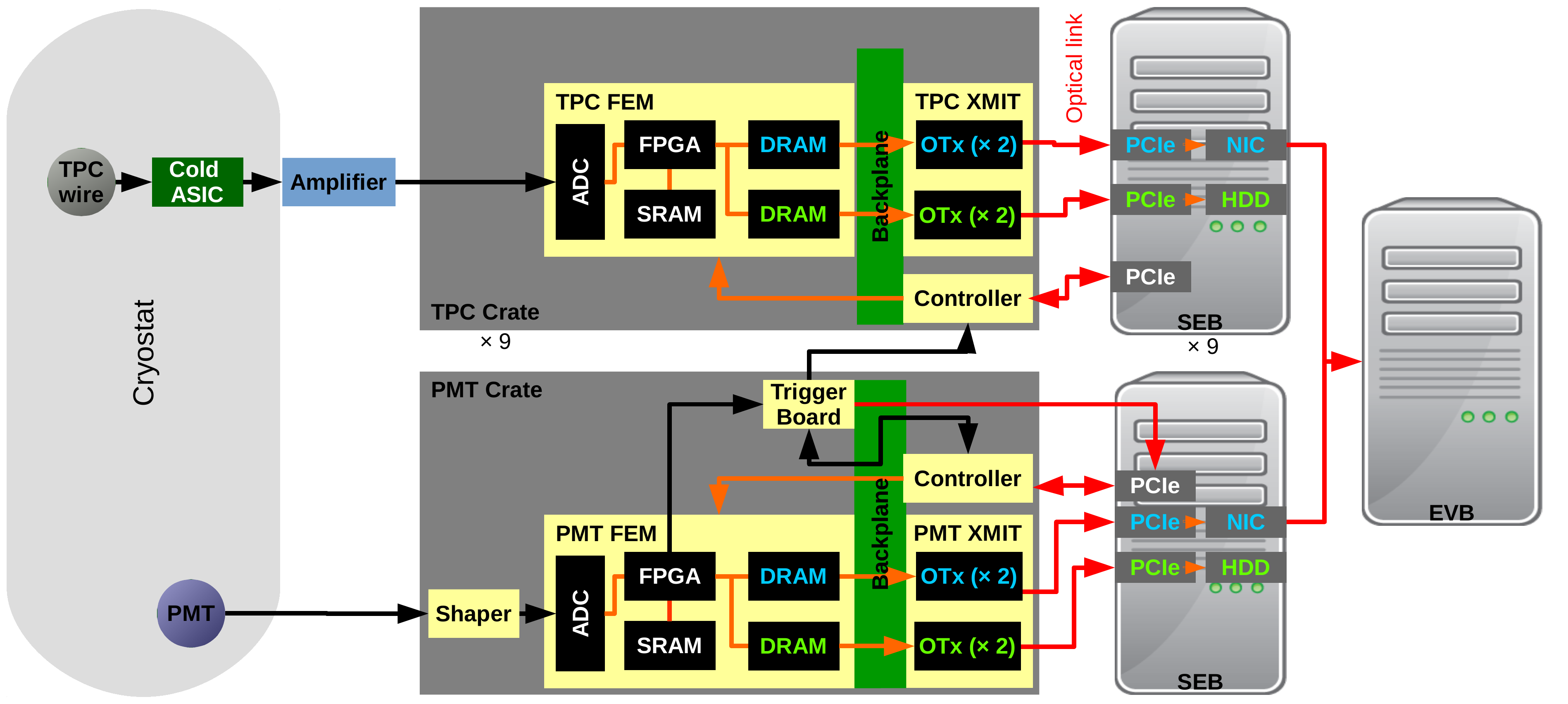}
\caption{Simplified diagram of the MicroBooNE readout dataflow. The trigger stream components are highlighted in blue text and the SN stream components are highlighted in light green.}
\label{fig:ReadoutElectronics}
\end{figure}

The MicroBooNE TPC is read out by three consecutive wire planes~\cite{Acciarri:2016smi}. The first two planes crossed by the drifting electrons are configured as induction planes; each having 2400 wires oriented at $\pm 60^\circ$ from the vertical. The last plane is configured as a collection plane, with 3456 vertical wires.
Figure~\ref{fig:ReadoutElectronics} illustrates the dataflow in the MicroBooNE readout, from a TPC wire to the DAQ server.
First, the signal in each TPC wire is preamplified and shaped by ASICs immersed in the liquid argon, and then extracted from the cryostat through feedthroughs. The signal is further amplified by warm electronics immediately upon extraction to condition it for transmission using shielded twisted-pair cables to the readout digital electronics on a platform above the detector.
The readout of the 8256 TPC channels is distributed among 9 readout crates, each one connected to a dedicated DAQ server, known as a sub-event buffer (SEB). Seven of these crates (labeled as crates~02 -- 08) are loaded with 15 front-end modules (FEMs), each reading out 64 channels, consisting of 16 wires from the first induction plane (plane U), 16 from the second induction plane (plane V) and 32 from the collection plane (plane Y). Crate~01 is loaded with 11 FEMs (one of them with only 32 channels) reading out exclusively induction channels from the first plane. Crate~09 is loaded with 14 FEMs (one of them with only 32 channels) and reads out mostly induction channels (720) from the second induction plane, with some channels (96) from the collection plane and some channels (48) from the first induction plane.
In each FEM, the TPC data is digitized by 8 octal-channel 12-bit~ADCs~\cite{AD9222} 
at $16~\rm{MS/s}$. An FPGA (Altera Stratix III~\cite{StratixIII}) downsamples the TPC data to $2~\rm{MS/s}$ and writes it in time order to a $1~\rm{M} \times 36~\rm{bit}$ $128~\rm{MHz}$ static RAM (SRAM) configured as a ring buffer. The TPC data is read back from the SRAM by the same FPGA, but now ordered by channel and split into two parallel streams. The trigger stream is only read out upon a trigger, generated by the trigger board and distributed through controllers to the FEMs. For a detailed description see~\cite{Acciarri:2016smi}. The SN stream is continuously read out. The data of each FEM is sent to a transmitter board (XMIT) through a readout crate backplane with bandwidth up to $512~\rm{MB/s}$. The backplane dataway is shared between both streams, with the trigger stream given priority over the SN stream using a token-passing scheme. The FEM has a dynamic RAM (DRAM) for each stream to buffer the data waiting for its turn to be transferred. The XMIT has 4 optical transceivers (OTx's), each rated to $3.125~\rm{Gb/s}$. Two are used for the trigger stream, and two for the SN stream. Finally, the data from each stream is read out by custom PCIe $1.0$ $\times 4$ cards in each SEB. 
The triggered data from each SEB is sent to the event builder (EVB) DAQ server using a network interface card (NIC), while the continuous stream is written to a $15~\rm{TB}$ local hard disk drive (HDD) in each SEB, awaiting a SNEWS alert to be further transferred to offline storage. If no SNEWS alert is issued and the disk occupancy reaches $80\%$, the oldest data is permanently deleted until the occupancy falls below $70\%$.

The large data rate read out by the front-end electronics in each readout crate, $\approx 4~\rm{GB/s}$, prevents the continuous acquisition of the TPC waveforms by the SEBs without compression. In particular, the bottleneck of the SN stream is found in the disk-writing speed of the local hard drive system, which is in the range of $50 - 200~\rm{MB/s}$. In order to achieve such data rates, the FPGA applies data reduction algorithms described in section~\ref{sec:DataReductionAlgorithms} to achieve a $\approx 20 - 80$ compression factor. 

The SN stream data is arranged into frames corresponding to $1.6~\rm{ms}$ of detector readout. Each FEM creates its own frame record, writing the TPC data in a payload which is preceded by a header consisting of twelve 16-bit words indicating the FEM address, a word count of the data in the payload, a sequential identifier, the frame number, and a simple checksum of the payload data. The TPC data consists of 16-bit words. The data from each channel is preceded by a channel header and a timestamp.

A continuous readout of the PMT system, based on the more conventional out-of-beam-spill discrimination signals described in~\cite{Acciarri:2016smi}, also exists but is not used for any of the results of this work.

\section{Data reduction algorithms}
\label{sec:DataReductionAlgorithms}
The SN stream FPGA firmware applies two data reduction algorithms sequentially. The first one is a ZS scheme applied individually to all the channels. With reference to an estimated baseline, the ADC samples are discarded if not meeting a configurable threshold.
The second data reduction algorithm is a fixed-table Huffman compression, in which consecutive ADC samples which differ by less than 4~ADC counts are encoded using a reduced number of bits.  

\subsection{Zero suppression}
\label{subsec:ZeroSuppression}
The ZS algorithm aims at removing the samples which do not carry any signal. For this, it checks whether an ADC sample passes an amplitude threshold after subtracting the channel baseline. The sign of the threshold can be chosen to be positive (passing samples are greater than the sum of the baseline and threshold), negative (passing samples are smaller than the baseline minus the threshold), or either. The threshold value and sign are configurable for each channel. The different methods used to determine the thresholds are described in section~\ref{sec:ZeroSuppressionThresholds} and appendix~\ref{app:PlanewiseThresholds}.
In addition, a number of samples preceding the first one to pass the threshold (presamples), and following the last sample that passed the threshold (postsamples) are retained in order to better capture the waveform. The set of samples that pass the threshold, plus the presamples and postsamples, is considered a Region-Of-Interest (ROI). The numbers of presamples and postsamples are configurable for each FEM (64-channel block). They have been set to 7 presamples and 8 postsamples, the maximum values allowed by the current FPGA firmware, to perform a local estimation of the baseline during offline analysis.

Two versions of the ZS firmware have been produced. One version uses a sliding-window algorithm in the FEM FPGA to estimate the baseline dynamically (see appendix~\ref{app:DynamicBaseline} for details). Another version uses a static baseline per channel which is configured at the beginning of the run. In the static baseline version, the baseline value for each channel is extracted from a previously chosen reference run from the DAQ trigger stream, taking the mode of the raw ADC distribution. 

\begin{figure}[htbp]
\centering
\includegraphics[width=1.0\textwidth]{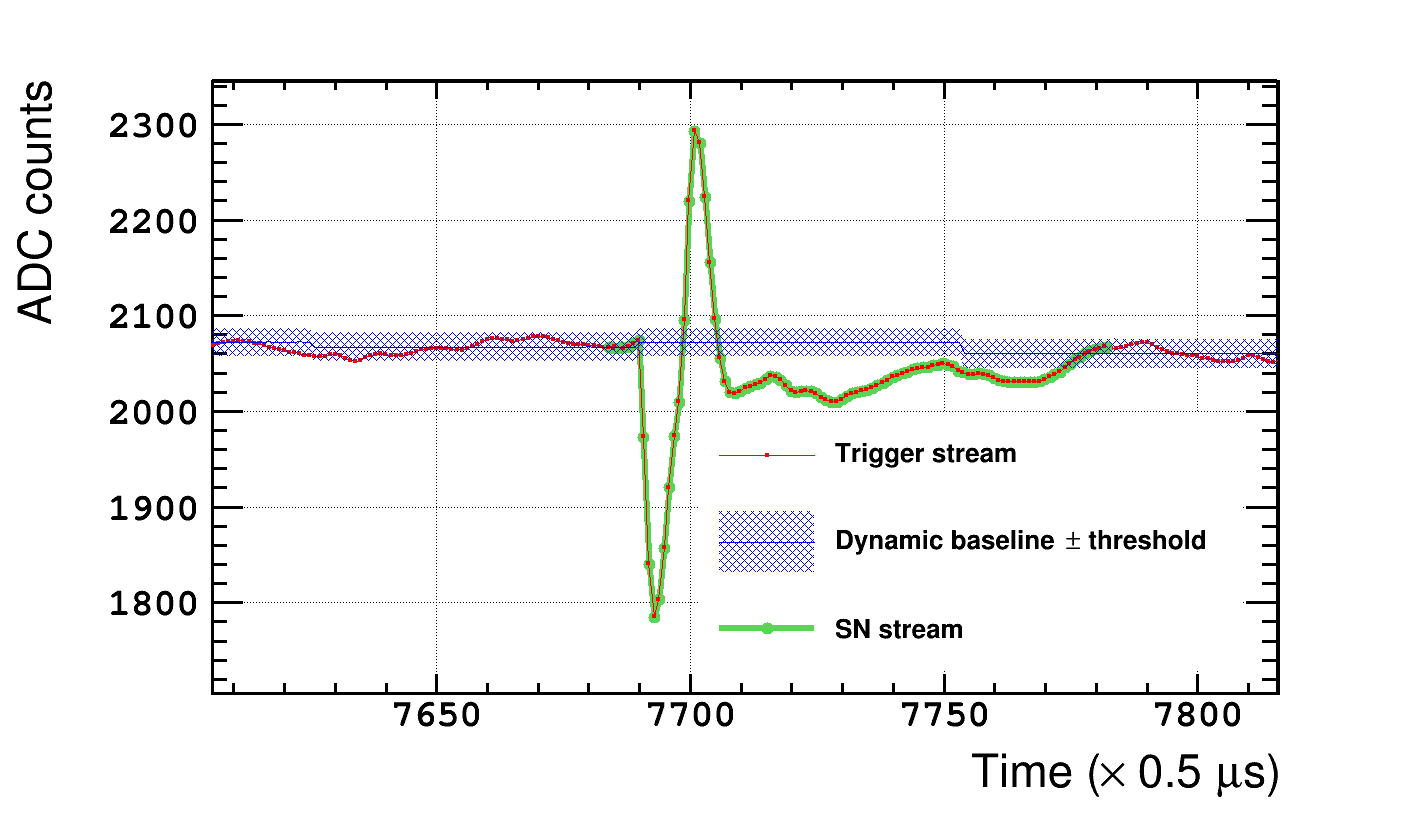}
\caption[Example of zero suppression.]{Example of data from a test stand at Nevis Laboratories showing the zero-suppressed waveform in the SN stream (green circles), overlaid on the same waveform from the trigger stream (red squares). An emulation of the dynamic baseline used by the FPGA is shown as a blue line, with the threshold shown as a blue cross-hatched band. Only the samples with their ADC values out of this band are saved, plus a number of samples preceding them (presamples) and following them (postsamples). The threshold and baseline estimation tolerance values used in this figure are for illustrative purposes and do not correspond to the ones used in the MicroBooNE detector.}
\label{fig:ZeroSuppressionExample}
\end{figure}

An illustration of the effect of ZS (with dynamic baseline) on a waveform is shown in figure~\ref{fig:ZeroSuppressionExample}.
The results from each configuration are further discussed in section~\ref{sec:DataRatesAndCompression}.

\subsection{Huffman encoding}
After zero suppression, the digitized waveform is run through a Huffman encoding~\cite{4051119} stage in which successive ADC samples differing by no more than $\pm3$~ADC counts relative to the predecessor ADC sample value are encoded as shown in table~\ref{tab:Huffman}.
This stage reduces the memory footprint of the saved waveform by attempting to store more ADC samples in the same memory space of a single uncompressed ADC sample.

\begin{table}[htbp]
\centering
\caption{Huffman encoding table relating the value of the difference between the current ADC sample and the preceding one, $\mathrm{\Delta ADC = ADC_i - ADC_{i-1}}$, and the Huffman binary code.}
\smallskip
\begin{tabular}{|r|r|}
\hline
$\mathrm{\Delta ADC}$ & Code\\
\hline
0 & 1\\
-1 & 01\\
+1 & 001\\
-2 & 0001\\
+2 & 00001\\
-3 & 000001\\
+3 & 0000001\\
\hline
\end{tabular}
\label{tab:Huffman}
\end{table}

The readout electronics data format uses 16-bit words. Non-Huffman-encoded ADC words use the lowest 12~bits to store the 12-bit ADC value, and use the rest as header to identify the word as non-Huffman encoded. Huffman-encoded words have the sixteenth bit (most significant bit) set to 1 to identify the word as Huffman-encoded. The other 15~bits are available to contain ADC information using the codes shown in table~\ref{tab:Huffman}.
Since Huffman-encoded ADC samples within the same 16-bit word need to be separated, the chosen Huffman encoding reserves the character 1 for punctuation.
If there are no more samples to be encoded in the Huffman word (because the next ADC difference is larger than $\pm3$~ADC counts) or the required code does not fit in the available bits, the unused least significant bits are filled with zeros. In the latter case, a new Huffman-encoded word will be created to continue storing the ADC differences.

\section{Configuration of zero suppression parameters}
\label{sec:ZeroSuppressionThresholds}
This section describes the improved method used since August 2018 to determine the channel-wise thresholds for ZS of TPC waveforms. Our initial approach, now deprecated, was to use a single physics-motivated threshold for each TPC plane (a plane-wide threshold) to separate signals from noise and is described in appendix~\ref{app:PlanewiseThresholds}. As will be discussed in section~\ref{sec:DataRatesAndCompression}, the initial thresholds were too high. This allowed the setting of lower thresholds not driven by the separation of signal from noise, but by exploiting the bandwidth of the readout electronics, as described next. The motivation is that any signal that is zero-suppressed online will be lost forever, while we can add additional higher thresholds offline to reject noise if needed.

\begin{figure}[htbp]
\centering
\includegraphics[width=0.8\textwidth]{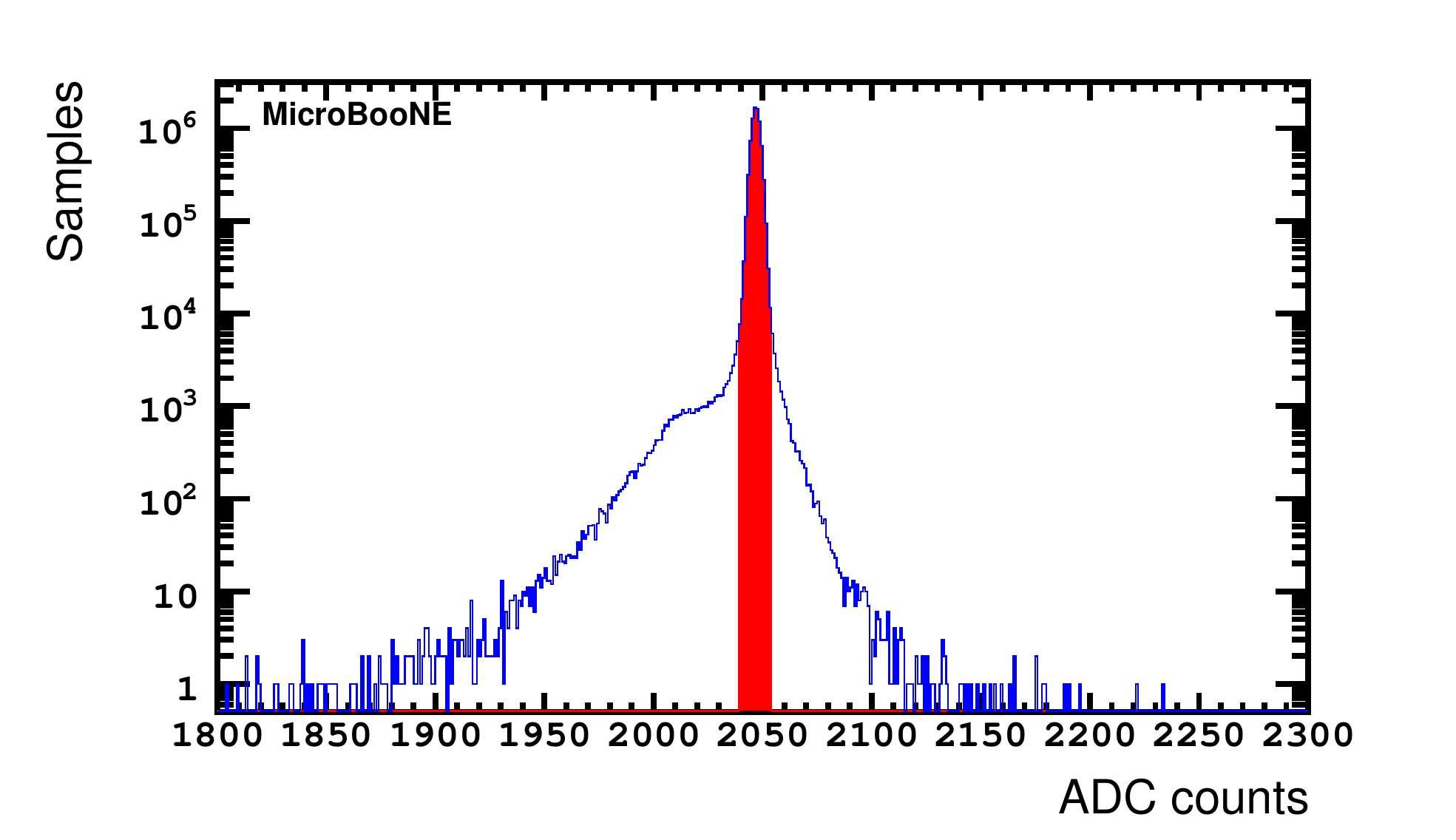}
\caption{Examples of raw ADC distributions from run 18468 from the trigger stream (taken during the 2018 beam shutdown) used to determine the ZS thresholds and static baseline for channel 2000 from the first induction plane. The red area shows the data that is zero-suppressed, found by integrating $98.5\%$ of the distribution symmetrically around the maximum (taken as the baseline value). The limits of the red area denote the position of the thresholds.}
\label{fig:RawADC18468}
\end{figure}

In order to set the threshold values per channel, we analyze the trigger stream ADC distribution for each channel. This distribution consists overwhelmingly of noise. We find the ADC values defining the shortest symmetrical interval around the mode of the ADC distribution, containing $98.5\%$ of the ADC distribution, corresponding to a ZS compression factor greater than 67.
An example of the method is shown in figure~\ref{fig:RawADC18468}.
The mode of the distribution is also taken as the baseline ADC value for the ZS firmware that uses static baselines as described in section~\ref{subsec:ZeroSuppression}.
The ADC values of the integration limits found, after subtracting the ADC baseline value, are taken as the ZS thresholds.

\begin{figure}[htbp]
\centering
\includegraphics[width=0.7\textwidth]{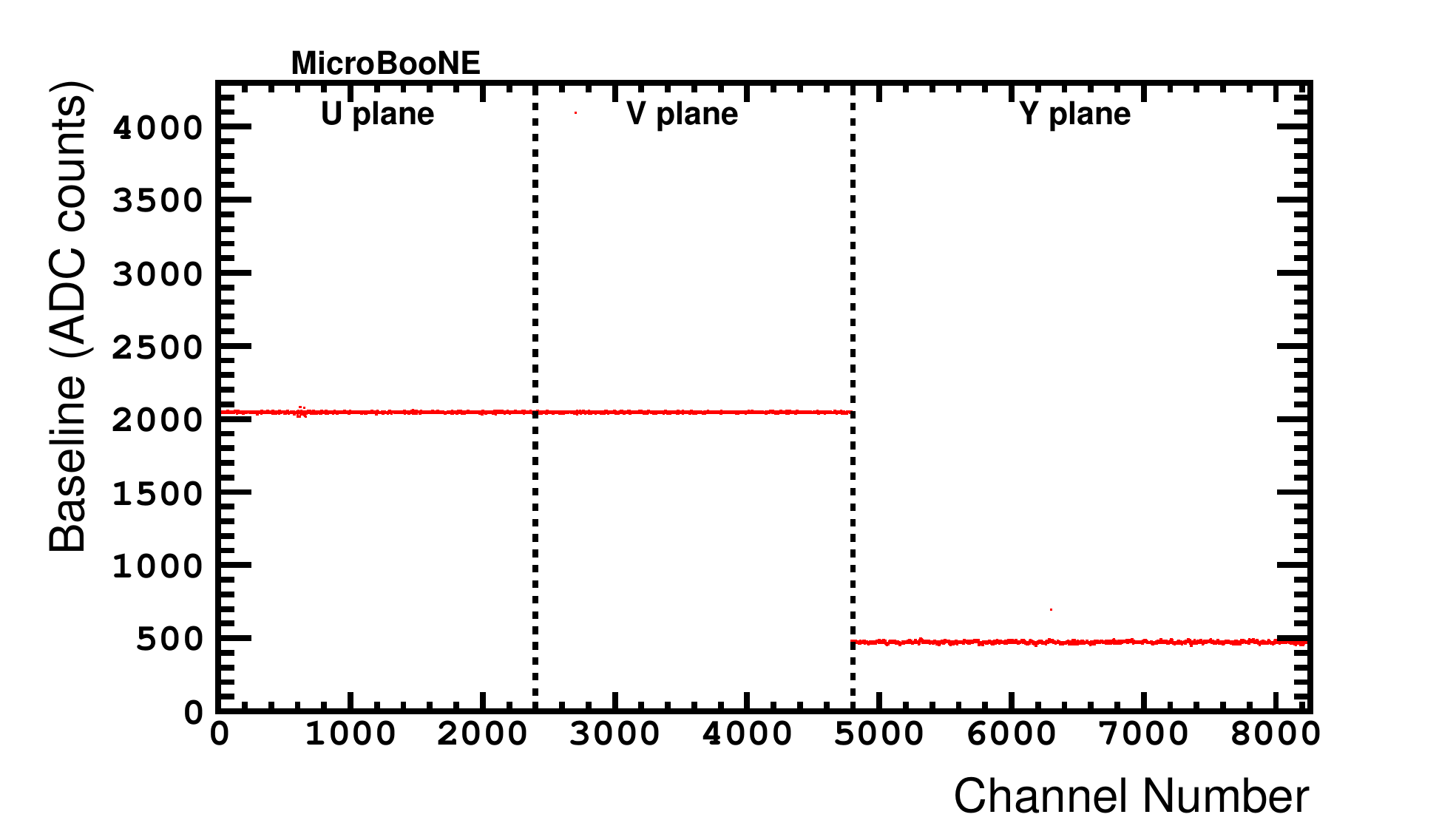}
\includegraphics[width=0.7\textwidth]{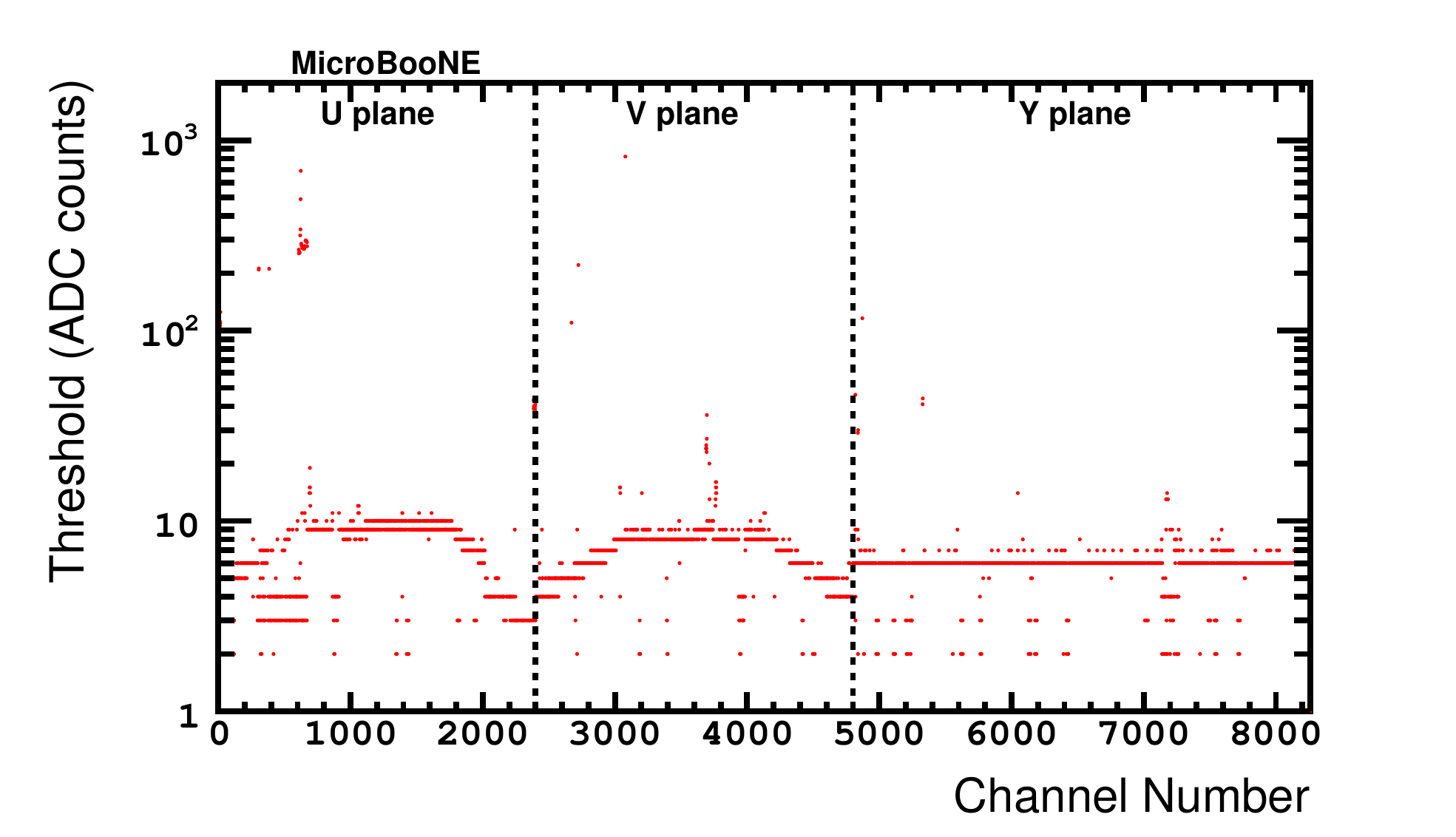} 
\caption[Channel-wise baselines and thresholds used for ZS]{Channel-wise baselines (top) and thresholds (applied bipolarly, bottom) used for ZS with the FPGA firmware with static baselines. The baseline values were set to optimize the dynamic range of each channel (bipolar signals on the U and V planes and unipolar positive signals on the Y plane). The overall threshold distribution follows the expected dependence on the wire length.}
\label{fig:Thresholds18468}
\end{figure}

The channel-wise thresholds and baselines used for the static-baseline firmware are shown in figure~\ref{fig:Thresholds18468}. 
Almost every channel has a threshold lower than the common plane-wide threshold counterpart.
The average threshold is 3.6~times smaller for U plane channels, 2.2~times smaller for V plane channels, and 5.2~times smaller for Y plane channels, allowing the recording of more data and ensuring a higher charge-collection efficiency for low-energy signals. Furthermore, noisy channels are effectively masked by setting higher thresholds for them.

\section{Data compression results}
\label{sec:DataRatesAndCompression}

\begin{figure}[htbp]
\centering
\includegraphics[width=1.0\textwidth]{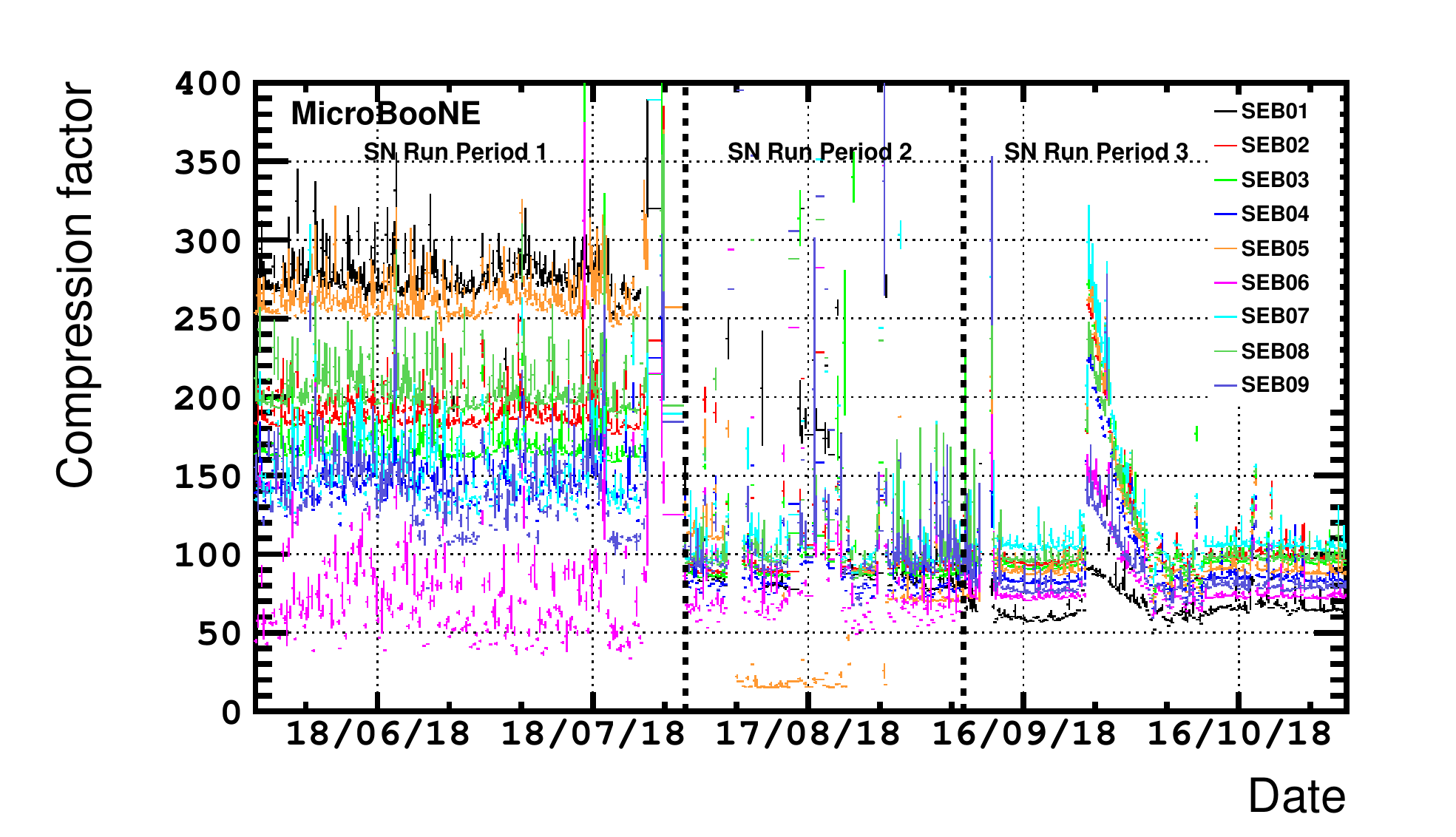} 
\caption[Compression factors achieved in the SN stream for the nine TPC DAQ servers with the three ZS configurations used so far]{Compression factors achieved in the SN stream for the 9 TPC DAQ servers (SEBs) with the three ZS configurations used so far. Date format is day/month/year. Until end of July 2018 (marked with a heavy dashed line) the configuration used the plane-wide thresholds and dynamic baselines. During August 2018, the lower channel-wise thresholds were tested, keeping the dynamic baseline. Beginning in September 2018 (marked with a heavy dashed line), the static baselines were introduced. Each point shows the mean of the compression factor over 6 hours. Error bars show the standard error on the mean. The low compression factor for SEB05 during mid-August was caused by a misconfigured front-end ASIC after emerging from one power outage, and returned to the proper configuration in the following power outage. The jump in compression factors starting on September $24^{\rm th}$, 2018 corresponds to the filling of the cryostat with a batch of lower-purity argon, followed by a period of high-voltage instabilities.}
\label{fig:CompressionFactorQuery}
\end{figure}

The compression factor is computed as the ratio of the expected data rate without compression to the measured compressed data rate. The SN stream data-dominant physics contribution is cosmic-ray muons crossing the MicroBooNE detector.
Figure~\ref{fig:CompressionFactorQuery} shows the compression factors achieved in each SEB for the three ZS configurations tested of the SN stream, summarized in table~\ref{tab:ZSConfigs}. 

\begin{table}[htbp]
\centering
\caption{Summary of ZS configurations tested in the MicroBooNE detector.}
\smallskip
\begin{tabular}{|c|c|c|}
\hline
\diagbox{Thresholds}{Baseline} & Dynamic & Static\\
\hline
Physics-driven plane-wide & SN Run Period 1& Not used\\
Bandwidth-driven channel-wise & SN Run Period 2& SN Run Period 3\\
\hline
\end{tabular}
\label{tab:ZSConfigs}
\end{table}

From November 2017 to July 2018 (SN Run Period 1), the ZS configuration described in appendix~\ref{app:PlanewiseThresholds} was used. 
This resulted in data rates well below the 50 MB/s target, except for SEB06. The cause of the high data rates and variation observed in SEB06 was traced back to the large number of noisy channels (due to ASIC misconfiguration) read out by that readout crate, which prevented the dynamic baseline algorithm from establishing an accurate and stable baseline upon which to execute the ZS. The dispersion in data rates is also seen in other SEBs (e.g.\ SEB07 and SEB09, but with a smaller magnitude). Moreover, the low data rates in the rest of the SEBs suggested it was actually feasible to lower the thresholds to gain efficiency for low-energy signals, even at the expense of recording some noise which could be eliminated during the offline reconstruction. This motivated deprecating the plane-wide thresholds in favor of (mostly lowered) individualized thresholds adjusted to produce data rates closer to the target goal.

Beginning in August 2018 (SN Run Period 2), the ZS per-channel thresholds described in section~\ref{sec:ZeroSuppressionThresholds} were deployed, keeping the dynamic baseline estimation. Because most of the thresholds 
were below the plane-wide values, this resulted in an increase of data rates for most of the SEBs, while the raising of thresholds for a few especially noisy channels significantly decreased the fluctuations in SEB06. The lack of an accurate and precise baseline during large portions of the run was a major concern. This motivated the replacement of the baseline estimation algorithm with the static configuration version, beginning in September 2018 (SN Run Period 3), in order to have a baseline value for ZS from the beginning of the run, regardless of the noise conditions.
As seen in figure \ref{fig:CompressionFactorQuery}, this latest ZS configuration using static baselines and channel-wise thresholds achieved the compression-factor target range and resulted in better stability for all the SEBs, and has been adopted as the default running mode.

\section{Analysis of Continuous Readout Stream data}
\label{sec:Analysis}
This section describes the offline analysis of the SN stream data using \path{LArSoft}~\cite{LArSoft, Snider:2017wjd} and \path{uboonecode}~\cite{uboonecode}, with the final goal of assessing the sensitivity to electrons with energies in the supernova neutrino range (few to tens of MeV).

\subsection{Event building}
\label{subsec:Assembler}
The binary data from the SN stream is written to files on each of the nine SEB DAQ servers, each of which has a $15~\rm{TB}$ RAID array, which temporarily stores the data for hours (typically more than $48~\rm{h}$) before it is permanently deleted. Data are processed for SN event building only on demand. Parallel processes independent of the primary DAQ run on each SEB, retrieve the requested frames from a given run number and send them via TCP (transmission control protocol) connections to a central server. Data from all TPC SEBs and the PMT SEB are assembled for each frame and written to a MicroBooNE-format file. Currently, there is no automated trigger to respond to the SNEWS alert given the large time window to react. Instead, the runs surrounding the alert timestamp would be marked as such by collaborators to prevent deletion in the hours following the alert, and subsequently assembled to search for supernova neutrinos.

The encapsulation of the SN stream is slightly more complex than the regular trigger stream since the data consists of ROIs occurring randomly in time. In the case of a core-collapse supernova, the neutrino burst is spread out over tens of seconds and there is no clear start time, so the event concept must be defined. For the Michel electron analysis described in section~\ref{subsec:ReconstructedMichelSpectrum}, the events were defined as 6400-samples long, the size used for the trigger stream. Each event is formed by taking a $1.6\rm{~ms}$-long frame (3200 samples), and appending the preceding last $0.8~\rm{ms}$ (1600 samples) from the previous frame and the following first $0.8~\rm{ms}$ (1600 samples) from the next frame. This ensures that objects crossing frame boundaries can be well reconstructed by the pattern recognition algorithms. For all the results from the SN stream shown in section~\ref{subsec:ReconstructedMichelSpectrum}, only the Michel electrons with the decay vertex in the central frame of each event are considered to avoid double-counting.

\subsection{Signal processing}
\label{subsec:BaselineSubtractionFlippedBitFilterDeconvolution}
\begin{figure}[htbp]
\centering
\includegraphics[width=0.5\textwidth]{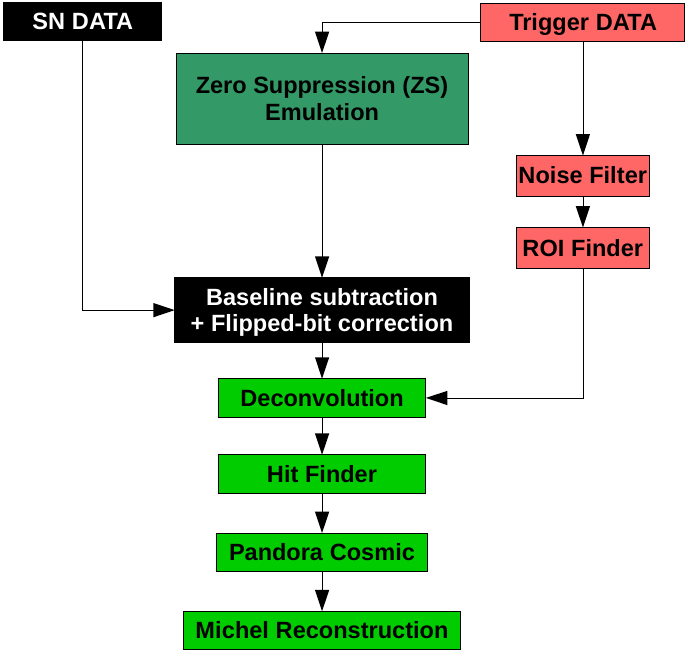}
\caption[Reconstruction chains.]{Illustration of the reconstruction chains used for the different data sets: the stages unique to the SN stream data are shown as black boxes, the stages unique to the trigger stream data are shown as red boxes. A ZS simulation stage emulates the real-time digital processing of the SN stream data, performed in the FPGA, and allows the trigger stream to be converted into continuous-stream-like data for direct comparisons. The light green boxes show common reconstruction stages for which the processing is identical.}
\label{fig:RecoChains}
\end{figure}

\begin{figure}[htbp]
\centering
\includegraphics[width=0.7\textwidth]{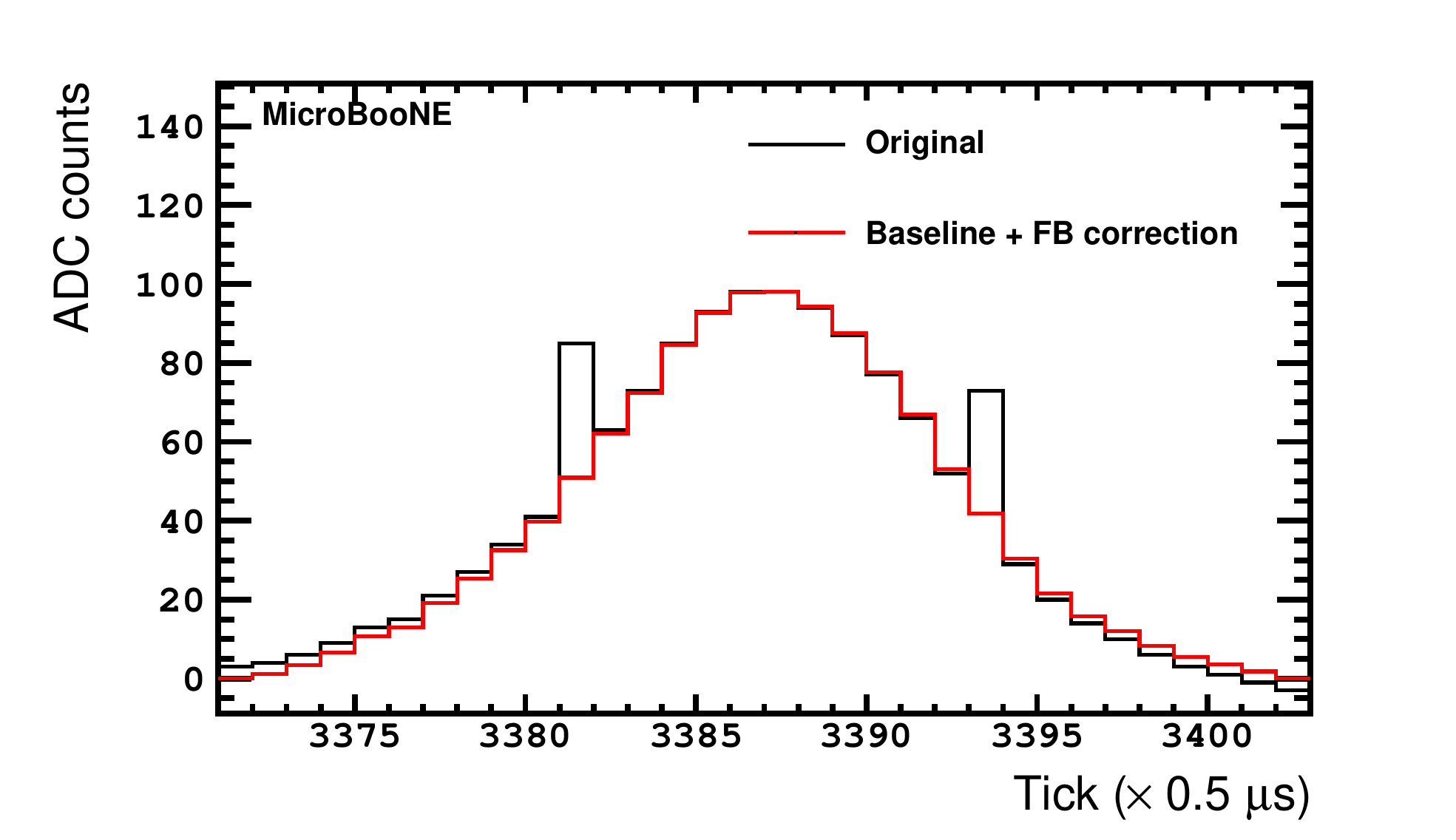}
\caption[Example of the flipped-bit filter.]{
Example of the flipped-bit filter on a waveform from a collection plane channel. The baseline-subtracted waveform after the flipped-bit filter is shown in red. The original waveform is shown in black (ADC counts have been shifted by the average baseline value to fit in the same axis range). ADC samples affected by flipped bits are seen at ticks~3381 and~3393 of the 2~MHz clock.}
\label{fig:FlippingBitFilter}
\end{figure}

The ROIs produced by the ZS in the SN stream follow a reconstruction chain similar to the one in the trigger stream (see figure~\ref{fig:RecoChains}). No noise-filter stage is used as the ZS removes the baseline regions where this filter is effective. No software ROI finder stage occurs either, since the ZS in the FPGA already produces ROIs. An additional challenge not present in the trigger stream is that $4\%$ of the ADC samples of the SN stream exhibit flipped bits randomly (see figure~\ref{fig:FlippingBitFilter}). This results in one or more of the bits encoding the ADC value to switch from 0 to 1, or vice versa, shifting the original ADC value by a combination of powers of 2. Since this has not been observed in the trigger stream, it must occur during digital processing after the SRAM, when the two streams separate (cf.~figure~\ref{fig:ReadoutElectronics}). While the investigation of the origin of the flipped bits continues, we have mitigated their effect offline as described next.

A tailored offline baseline subtraction is implemented for the SN stream using a linear interpolation between one of the presamples and one of the postsamples which are acquired during the ZS. Due to the occurrence of the flipped bits, the ADC value of the presamples and postsamples needs to be checked before using it for interpolation. For this, we compute the median ADC value for the presamples and the postsamples separately. 
This provides a first estimation of the baseline on each side of the pulse, since most of the samples are not affected by flipped bits.
Then we compare the presamples (postsamples) to the median and choose the earliest presample (latest postsample) within 15~ADC counts (absolute difference) 
of the median as the reference points for interpolation. The $15~\rm{ADC}$ counts cut was chosen to reject samples affected by a flipped bit in the fourth bit or higher, as that is the minimum deviation which could be identified as flipped bits as opposed to noise fluctuations or ionization charge signal-related deviation.

Once the baseline has been subtracted, the waveform goes through a filter to correct for flipped bits. This is done by comparing each ADC value to a linear interpolation built using the preceding and the following samples.
If the difference between them is larger than $32~\rm{ADC}$ counts, the ADC value is replaced with the interpolated value. Flipping bits with a shift smaller than $32~\rm{ADC}$ counts are difficult to distinguish from actual signals and we do not attempt to correct them. The ADC cuts were chosen by hand-scanning waveforms and identifying the spikes which could be attributed to flipped bits. Unlike in the case of the baseline estimation, identifying flipped bits in the rising or falling edge of a pulse is more challenging, which motivated the choice of more conservative values than for the baseline selection. An example of this algorithm is shown in figure~\ref{fig:FlippingBitFilter}.

Finally, the waveform is processed using the same 1-D deconvolution tool~\cite{Adams:2018dra, Adams:2018gbi} that is used to deconvolve the trigger stream waveforms.

\subsection{Michel electron reconstruction}
\label{subsec:ReconstructedMichelSpectrum}
In order to show the performance of the SN stream, we use Michel electrons from stopping cosmic-ray muons. The Michel electron spectrum spans an energy range very similar to the electrons resulting from charged-current interactions of electron neutrinos from a core-collapse supernova. The reconstruction and selection follows the MicroBooNE Michel electron analysis~\cite{Acciarri:2017sjy}, but extends its application to the induction planes. For each plane, the Michel reconstruction is processed independently; this allows the study of the effect of ZS on induction (bipolar) signals compared to collection (unipolar) signals.

The SN stream data set uses 1999022 frames, corresponding to 53.31 minutes of data, taken on September $21^{\rm st}$, 2018. To provide a reference for comparison, we use 1102845 events from off-beam zero-bias triggers from the trigger stream, corresponding to 58.82 minutes of data, taken between December $1^{\rm st}$, 2017 and July $7^{\rm th}$, 2018, after applying data quality criteria for the detector operating conditions. This data set is processed following the standard reconstruction for the trigger stream. In addition, we process the trigger stream data set through a ZS emulation that reproduces the FPGA algorithm and reconstruct it with the same tools as the SN stream.

\begin{figure}[htbp]
\centering
        \subfigure[Y plane (standard trigger stream).]{
            \includegraphics[width=.47\linewidth]{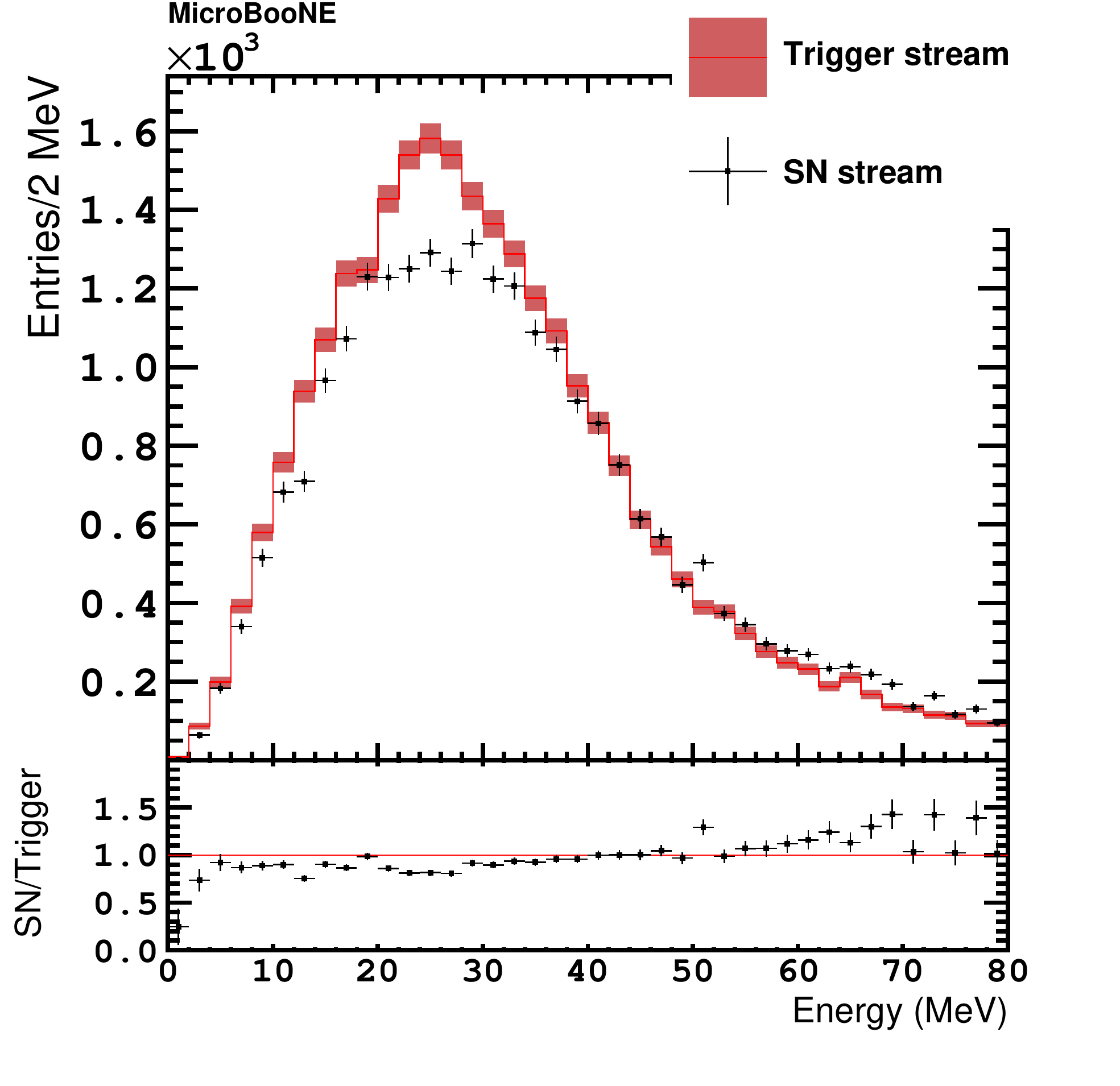}
            \label{fig:MichelTotalSpectrum2}
        }
        \subfigure[Y plane (trigger stream with ZS emulation).]{
           \includegraphics[width=.47\linewidth]{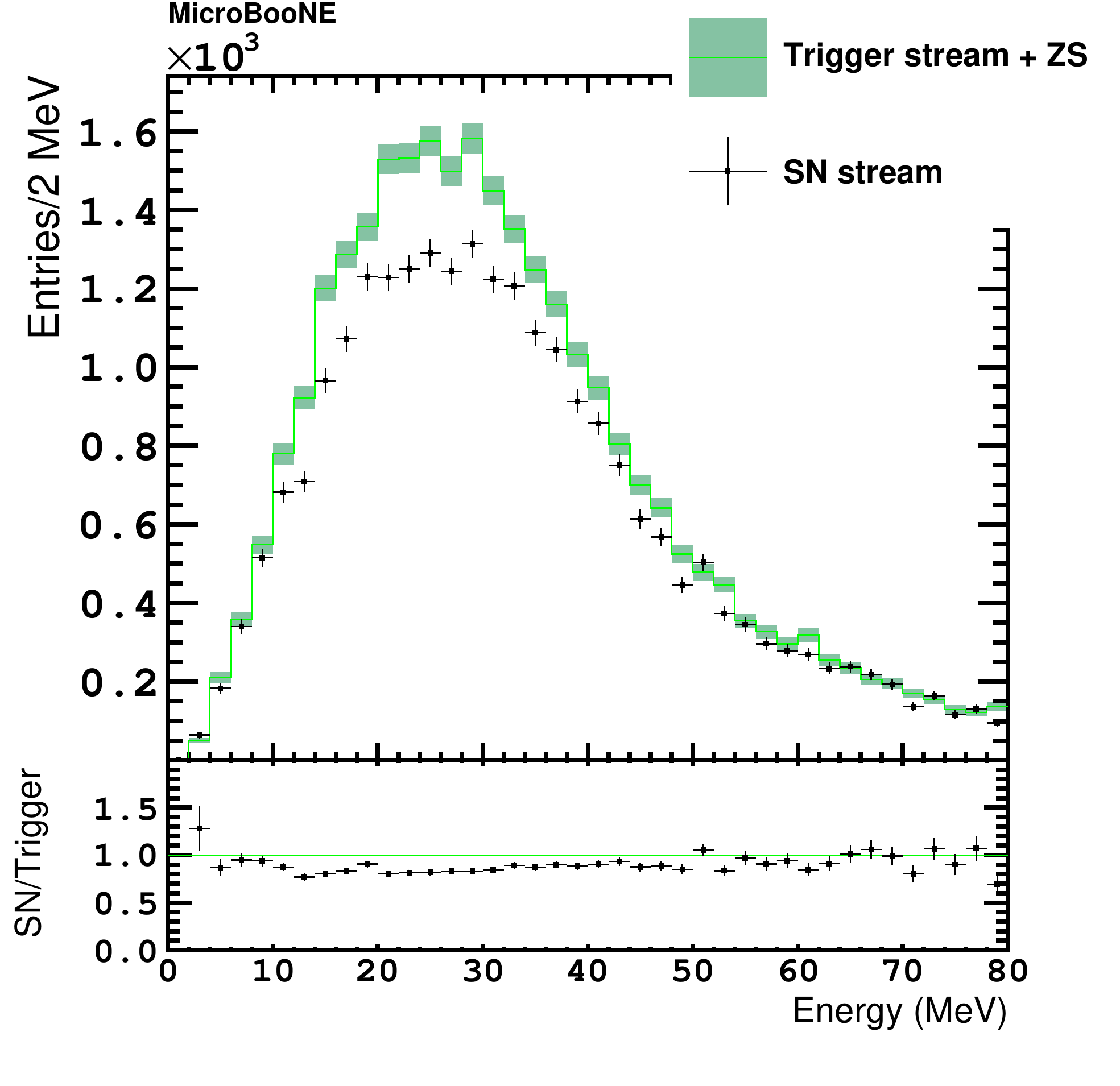}
           \label{fig:ZSMichelTotalSpectrum2}
        }
	  \caption[Michel electron candidate energy spectra on the collection plane.]{Michel electron candidate energy spectra reconstructed using only the collection plane. The black points in the upper panels show the same spectra from the SN stream. \subref{fig:MichelTotalSpectrum2} shows the trigger stream data processed through the standard reconstruction (red histogram). \subref{fig:ZSMichelTotalSpectrum2} shows the trigger stream data processed through the ZS emulation and the continuous stream reconstruction (green histogram). Both trigger stream spectra are normalized to the exposure of the SN stream. The bottom panels show the ratio between the SN stream and the trigger stream data points. The error bars and bands show statistical uncertainties.
	  }
	\label{fig:MichelTotalSpectra2}
\end{figure}

\begin{figure}[htbp]
\centering
        \subfigure[U plane (standard trigger stream).]{
            \includegraphics[width=.47\linewidth]{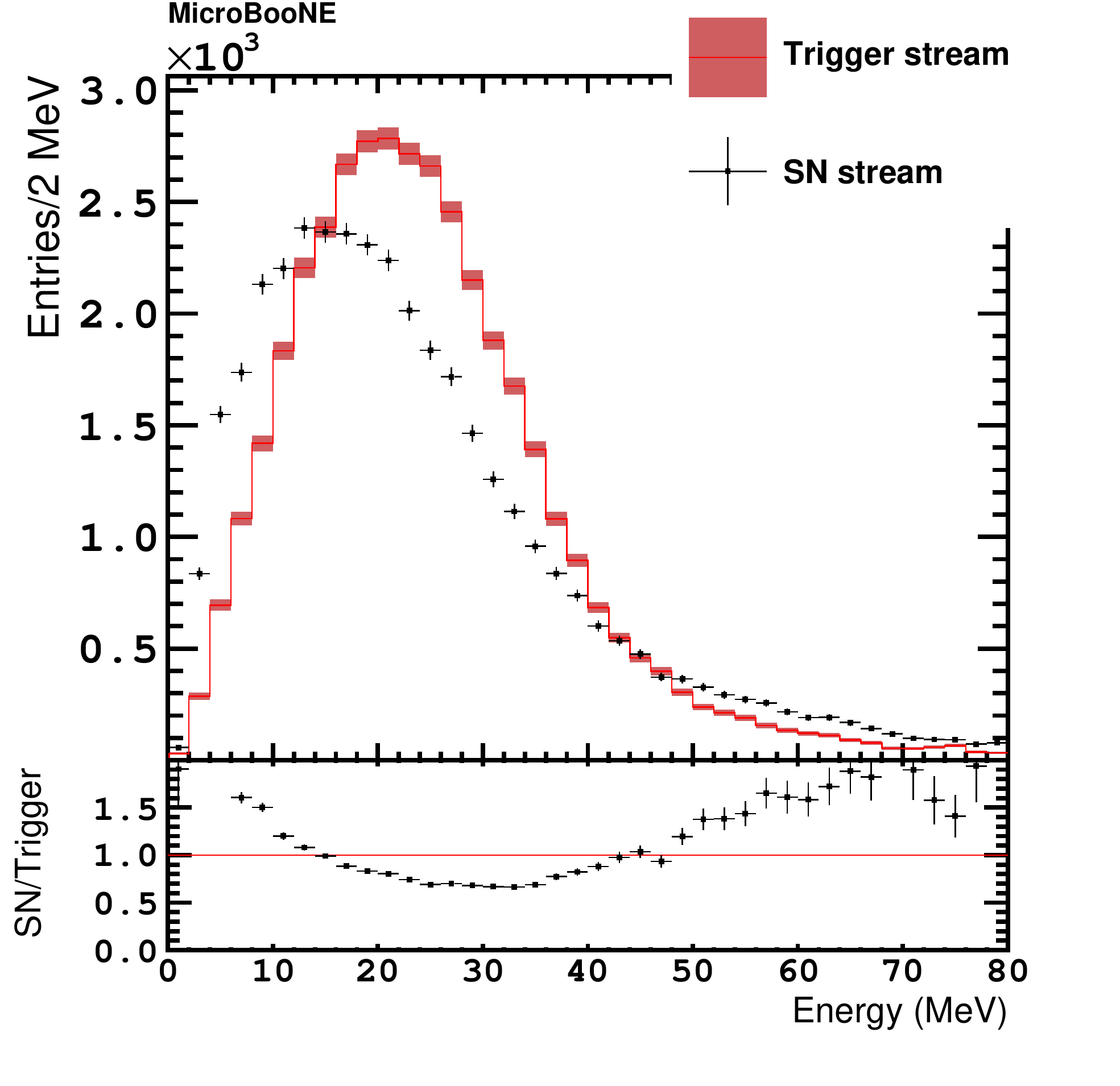}
            \label{fig:MichelTotalSpectrum0}
        }
        \subfigure[U plane (trigger stream with ZS emulation).]{
           \includegraphics[width=.47\linewidth]{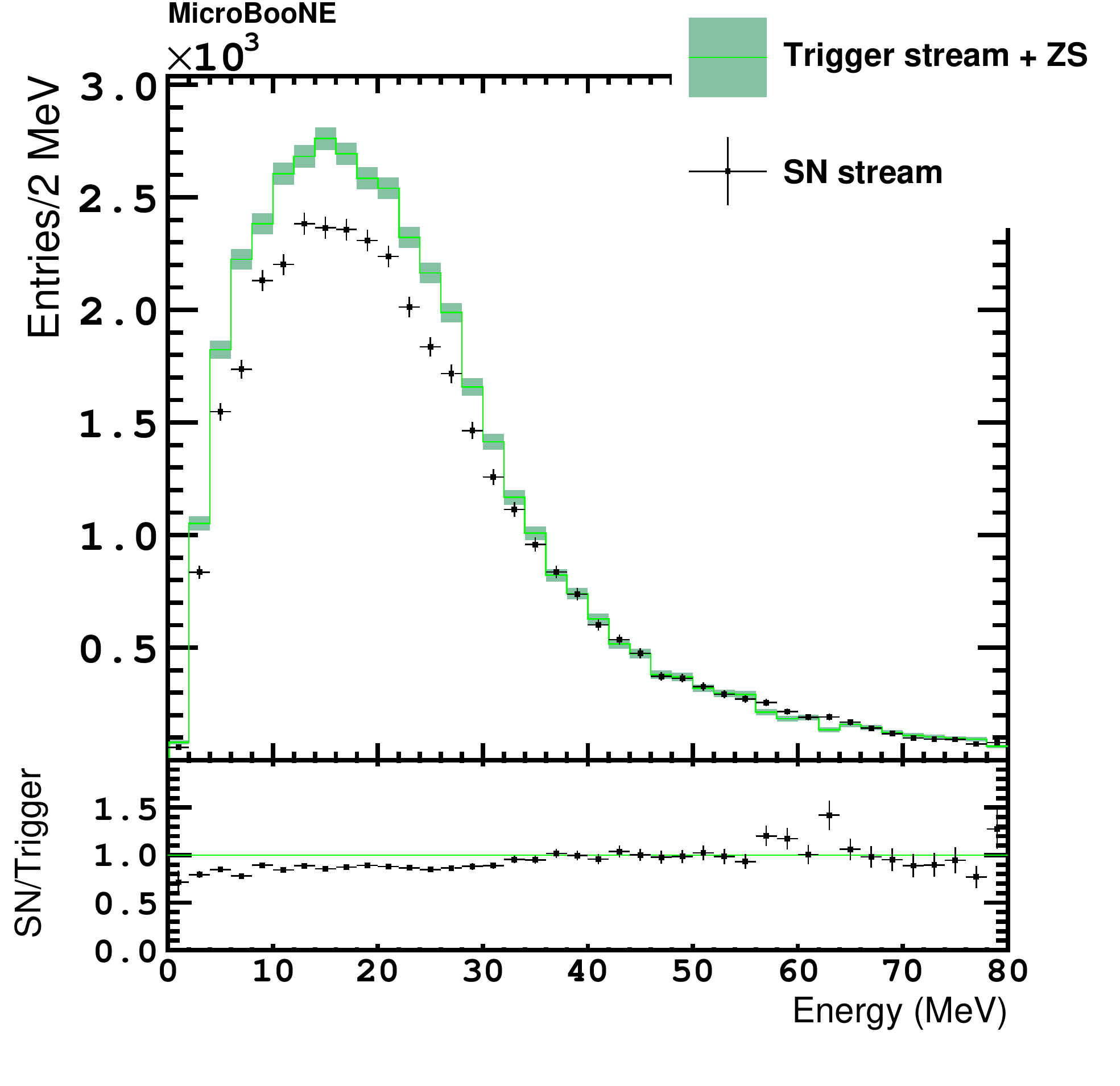}
            \label{fig:ZSMichelTotalSpectrum0}
        }
        \subfigure[V plane (standard trigger stream).]{
            \includegraphics[width=.47\linewidth]{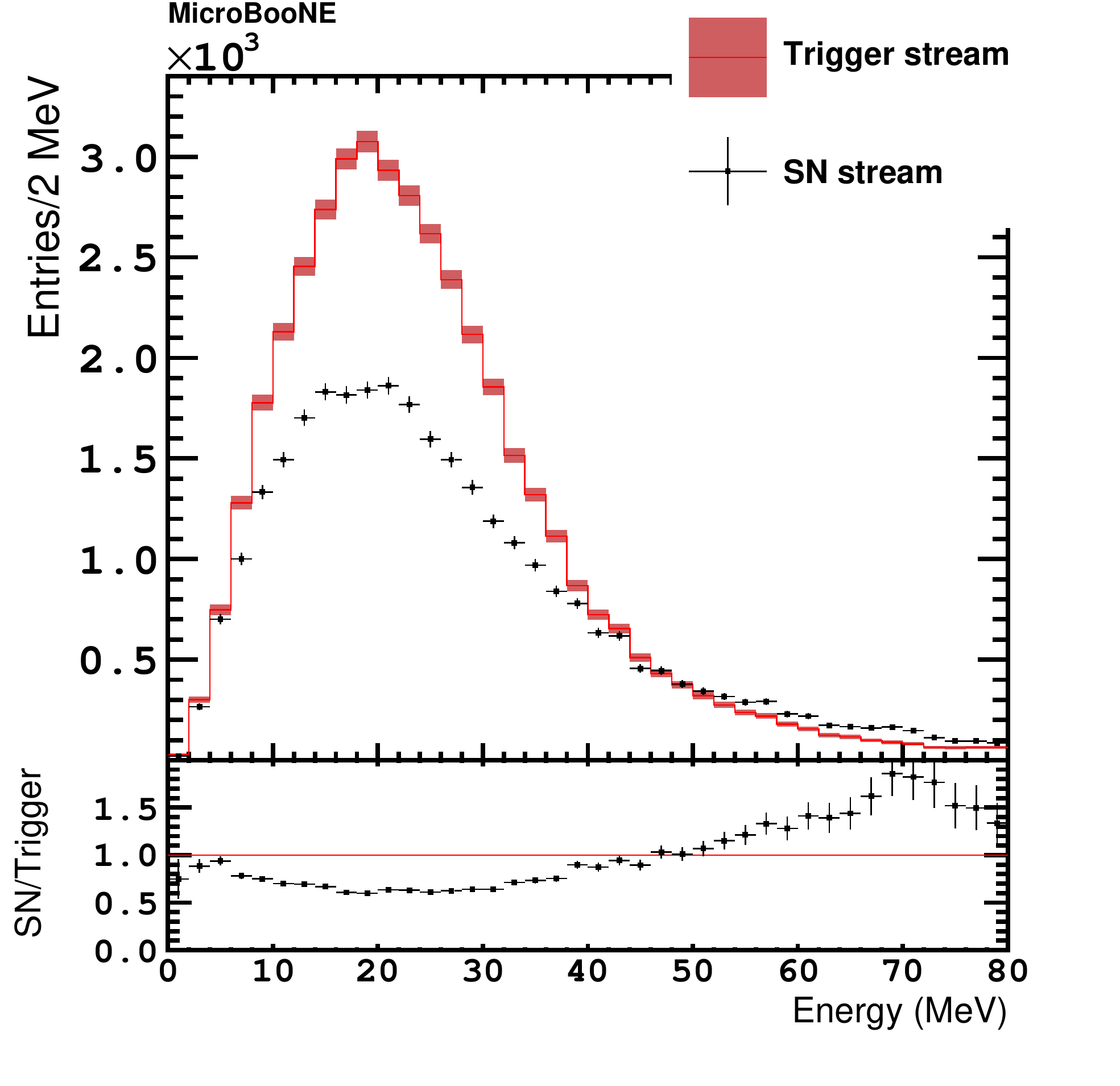}
            \label{fig:MichelTotalSpectrum1}
        }
        \subfigure[V plane (trigger stream with ZS emulation).]{
           \includegraphics[width=.47\linewidth]{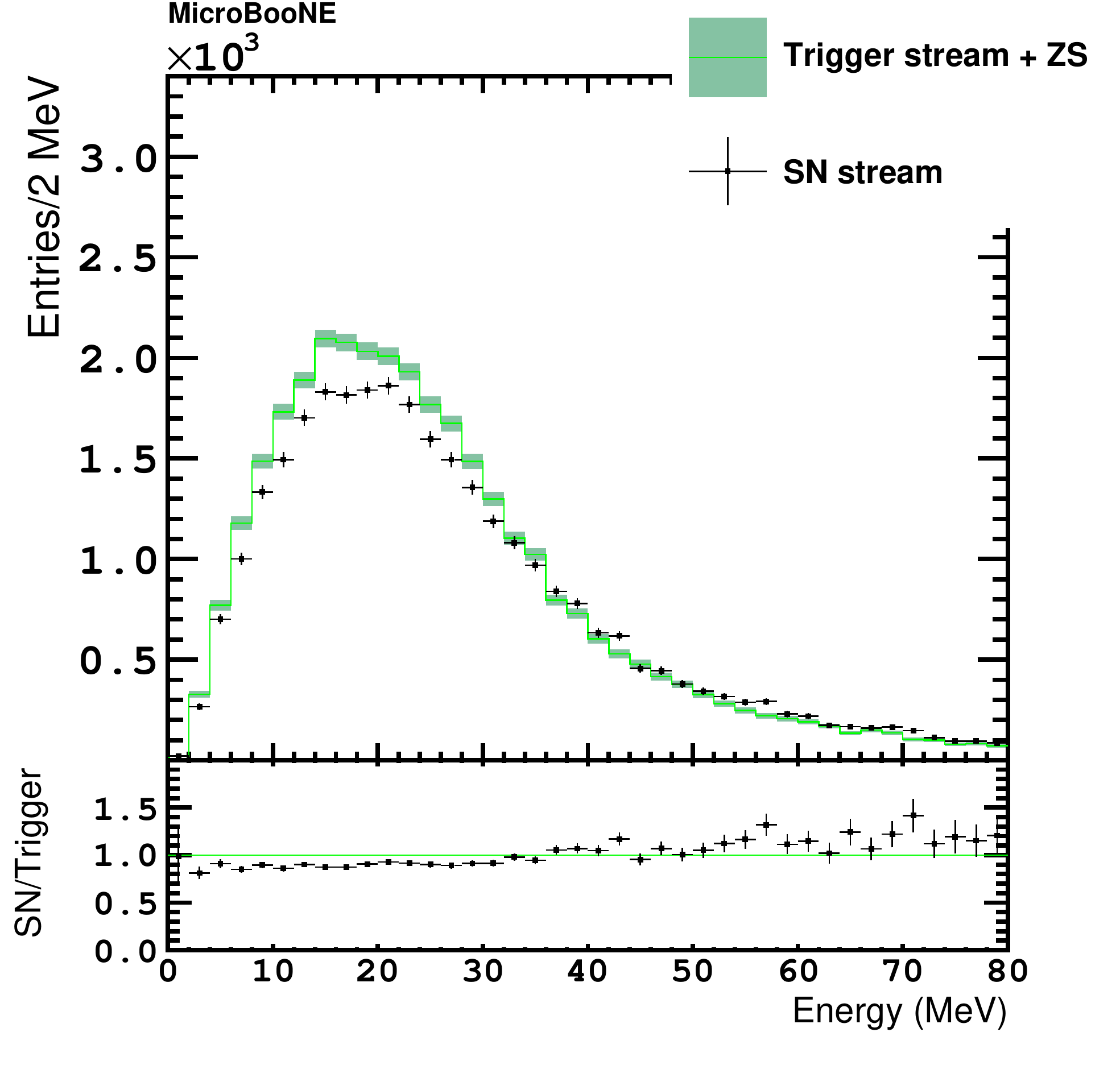}
            \label{fig:ZSMichelTotalSpectrum1}
        }
	  \caption[Michel electron candidate energy spectra on the induction planes.]{Michel electron candidate energy spectra reconstructed using only one of the induction planes. The top row shows the first induction plane (plane U), the bottom row shows the second induction plane (plane V). For each row, the markers and colors follow the same convention as figure~\ref{fig:MichelTotalSpectra2}.
	  }
	\label{fig:MichelTotalSpectraInduction}
\end{figure}

\begin{figure}[htbp]
\centering
\subfigure[Zero-suppressed raw waveforms.]{
  \includegraphics[clip, trim=1.4cm 9cm 2.5cm 8.20cm, width=0.475\textwidth]{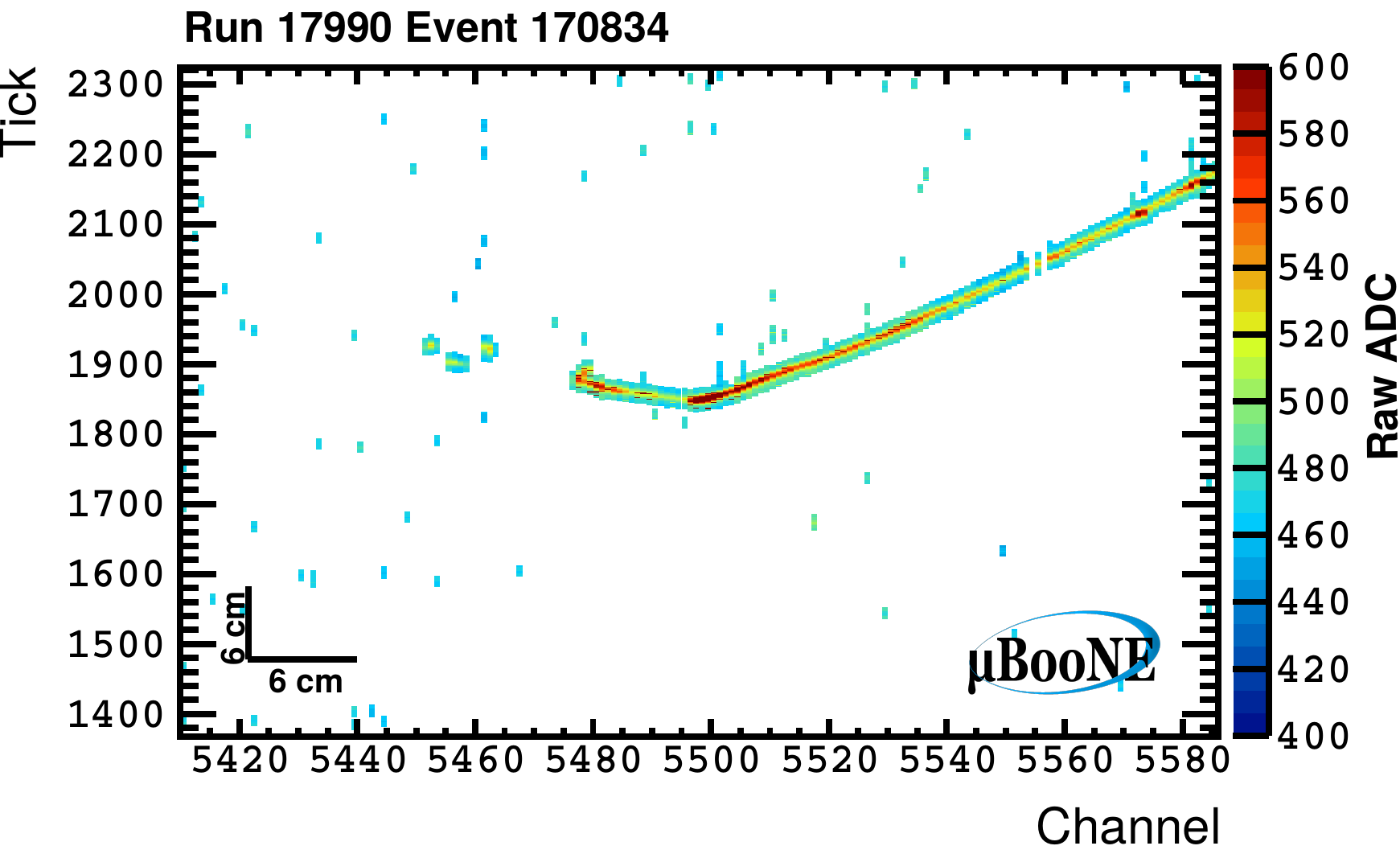}
  \label{fig:evdMichelHitsBoxes_raw}
}
\subfigure[Reconstructed hits.]{
  \includegraphics[clip, trim=1.4cm 9cm 2.5cm 8.20cm, width=0.475\textwidth]{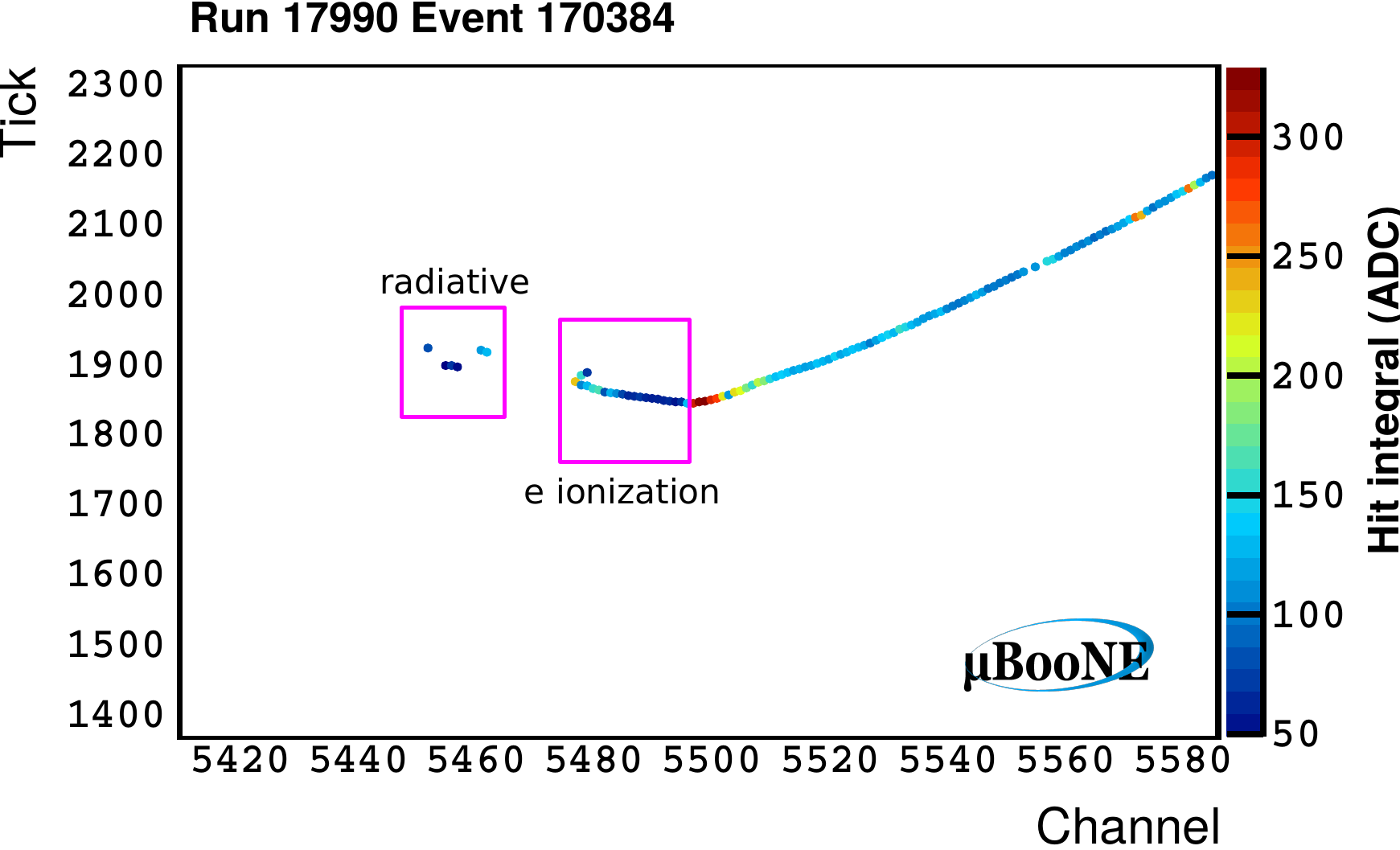}
  \label{fig:evdMichelHitsBoxes_hits}
}
\caption[Michel electron candidate event display.]{Michel electron candidate event display on the collection plane. The white background areas on \subref{fig:evdMichelHitsBoxes_raw} show the channel readouts which have been zero suppressed. The pink boxes on \subref{fig:evdMichelHitsBoxes_hits} illustrate the electron ionization and radiative components.}
\label{fig:evdMichelHitsBoxes}
\end{figure}

\begin{figure}[htbp]
\centering
\subfigure[U plane.]{
  \includegraphics[width=0.47\textwidth]{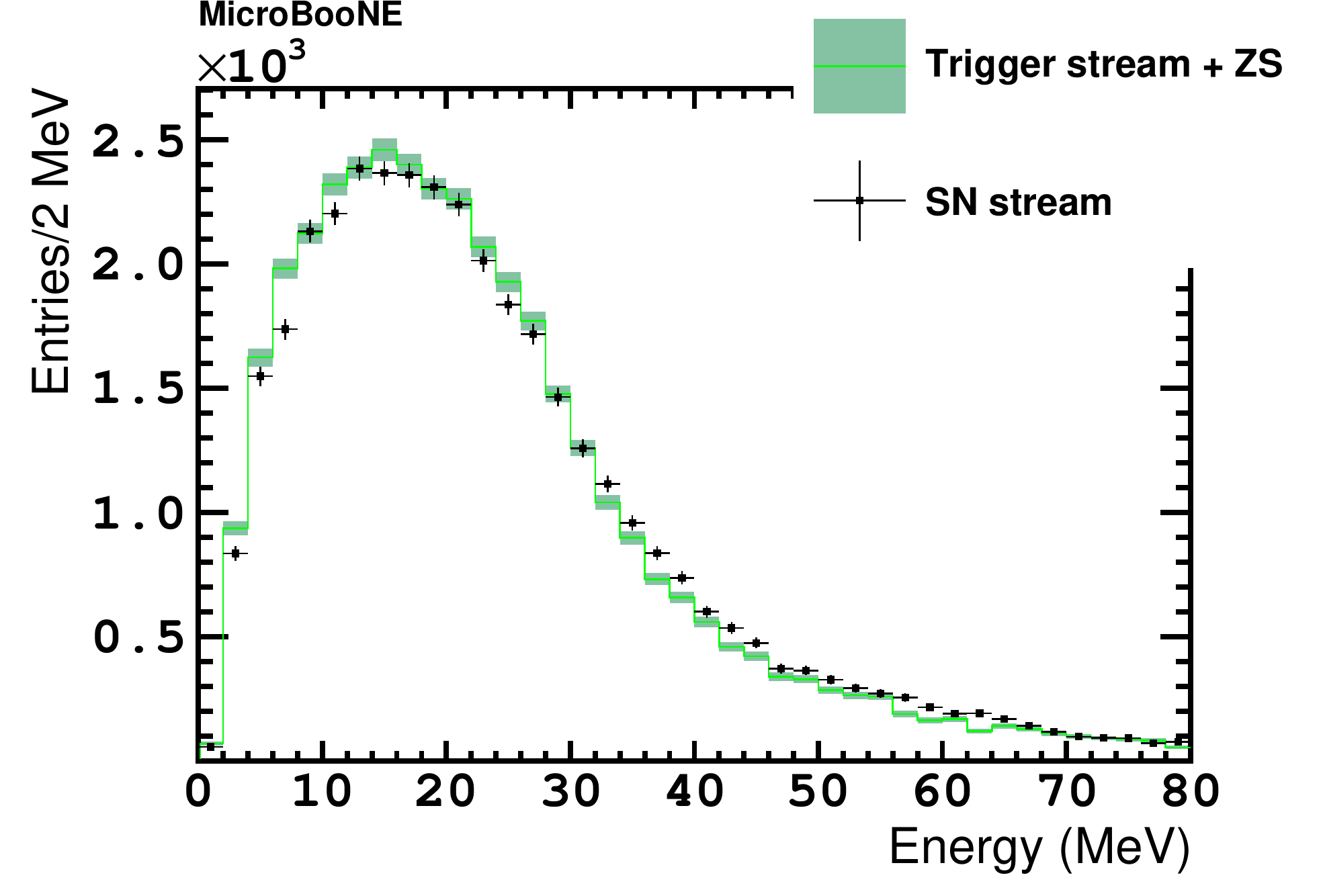}
  \label{fig:RelNormZSMichelTotalSpectra0}
}
\subfigure[V plane.]{
  \includegraphics[width=0.47\textwidth]{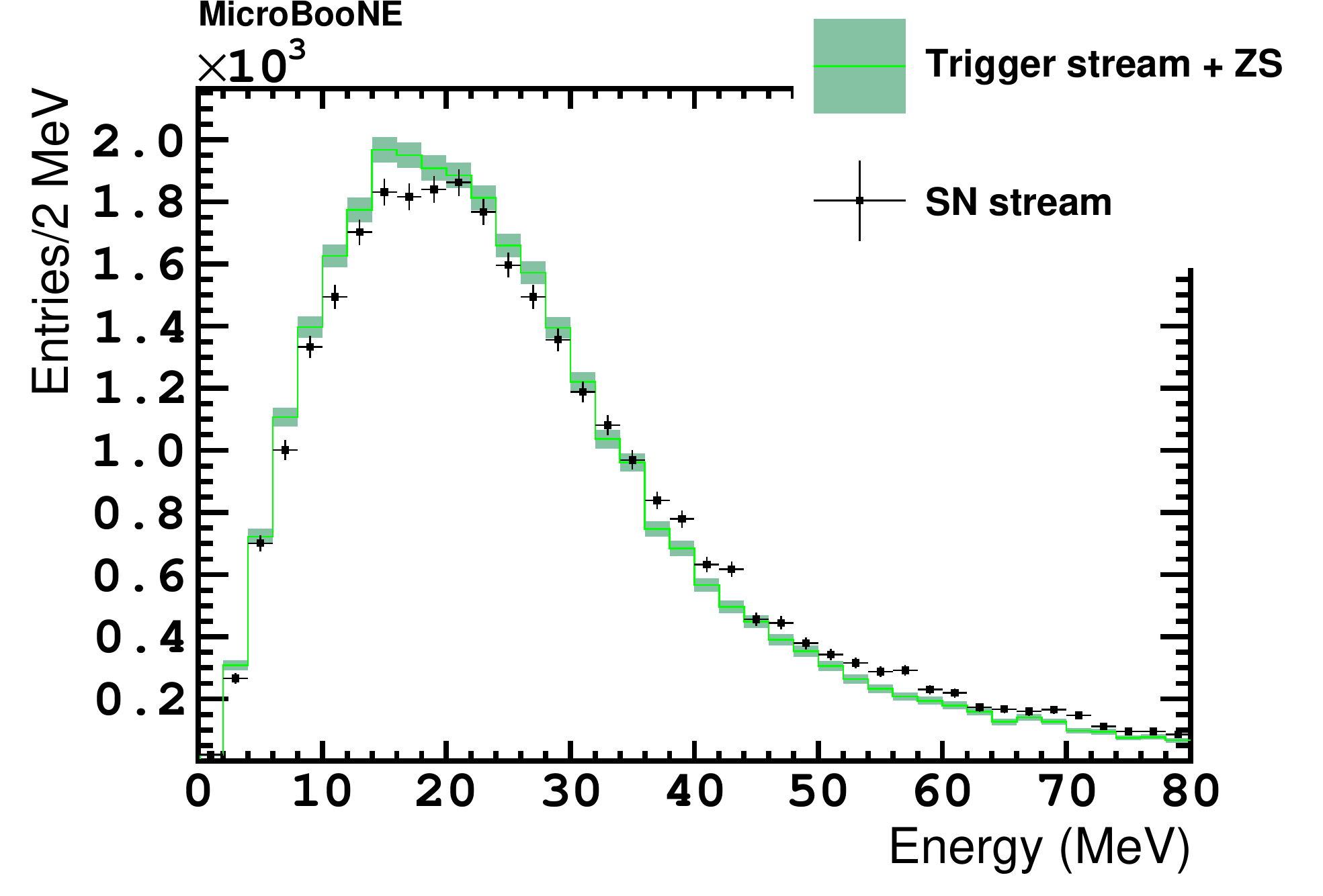}
  \label{fig:RelNormZSMichelTotalSpectra1}
}
\subfigure[Y plane.]{
  \includegraphics[width=0.47\textwidth]{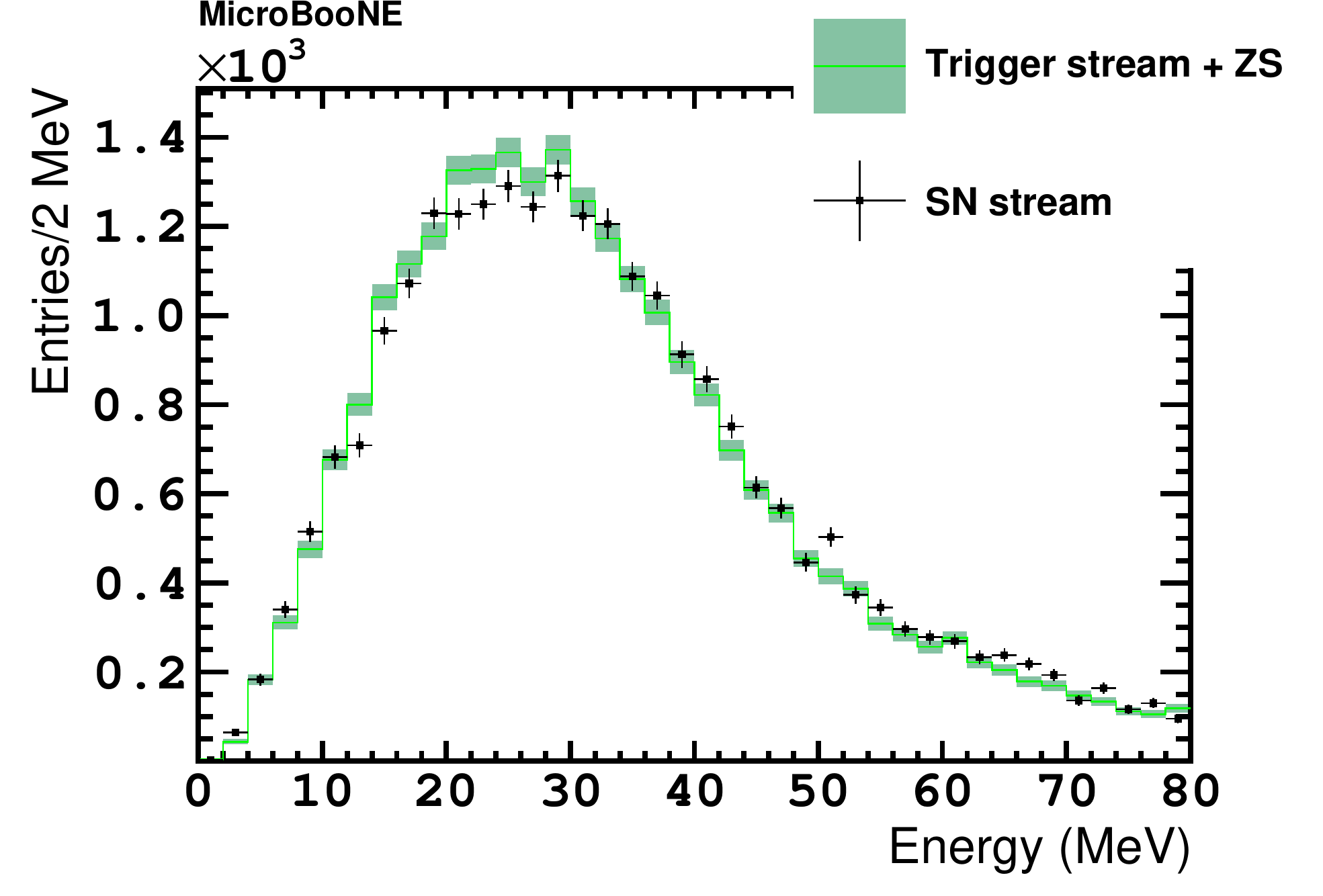}
  \label{fig:RelNormZSMichelTotalSpectra2}
}
\caption[Relatively normalized Michel electron candidate energy spectra.]{Michel electron candidate energy spectra from the SN stream (black points) overlaid on reference spectra from the trigger stream with simulated ZS (green histogram) normalized to the same area. Error bars and bands display statistical uncertainty.}
\label{fig:RelNormZSMichelTotalSpectra}
\end{figure}

The Michel electron spectra from the three TPC planes are shown in figures~\ref{fig:MichelTotalSpectra2} and~\ref{fig:MichelTotalSpectraInduction}. They show the total energy of the Michel electron candidates, summing over all the hits of the ionization and radiative clusters as in reference~\cite{Acciarri:2017sjy} (see an example in figure~\ref{fig:evdMichelHitsBoxes}). In order to avoid ambiguities when reconstructing Michel electrons with overlapping radiative components, we reject events with more than one Michel electron candidate. The ADC count-to-MeV calibration constants for the U, V and Y planes are $9.00 \times 10^{-3}~\rm{MeV/ADC}$, $8.71 \times 10^{-3}~\rm{MeV/ADC}$ and $9.24 \times 10^{-3}~\rm{MeV/ADC}$, respectively~\cite{Adams:2019ssg}. In the three figures, a rate and shape discrepancy between the SN stream and the trigger stream is observed. The shape discrepancy is caused by the ZS, as evidenced by a better agreement (i.e.\ flat ratio) with the trigger stream when we simulate the ZS (see the shape comparison in figure~\ref{fig:RelNormZSMichelTotalSpectra}).

Table~\ref{tab:MichelRates} shows the Michel electron rates, considering only the candidates with energy below $80~\rm{MeV}$, as in~\cite{Acciarri:2017sjy}. The rate and shape discrepancy in the second induction plane (plane V) when compared to the non zero-suppressed trigger stream is especially remarkable. The second induction plane has signals which are more vulnerable to not passing the ZS due to their smaller amplitudes and bipolar symmetric shapes, that favors cancellations. Excluding this plane, table~\ref{tab:MichelRates} shows that the Michel electron rates of the SN stream and the trigger stream agree within $10\%$. It is important to take into account that the SN stream data set corresponds to 53.31 minutes of actual run time while the trigger stream data set is spread over 218 days of run time, and hence are subject to different fluctuations. In particular, for the trigger stream data set, the average ground temperature was $5.6$ degrees Celsius and the average barometric pressure was $1017~\rm{mbar}$, while the average temperature and pressure for the SN stream data set was $26.6$ degrees Celsius and $1009~\rm{mbar}$, respectively. It is known that the seasonal temperature and pressure variations induce a modulation on the cosmic muon ray flux, as they change the density of the atmosphere in which the muons are produced. Determining the exact effect on the stopping muons inside the MicroBooNE detector is out of the scope of this work, but variations up to $20\%$ have been observed~\cite{Bernero:2013rva}. The slight increase in the trigger stream rates when simulating ZS over the trigger stream, and the associated decrease in the ratios of the rates (except for the V plane discussed above), is understood as events from the overflow bin (above $80~\rm{MeV}$, not included in the rate measurement) migrating into lower energies when adding the effect of ZS.

\begin{table}
\centering
\caption{Michel electron candidate rates measured in the SN stream (SN), the trigger stream (Trigger) and the trigger stream with simulated ZS (Trigger + ZS) on the three TPC planes. The last two rows show the ratio between the continuous stream and the trigger stream (without and with simulated ZS). Uncertainties are statistical only.}
\label{tab:MichelRates}
\smallskip
\begin{tabular}{|c|c|c|c|}
\hline
Michel e rate& U plane & V plane & Y plane\\
\hline
SN ($\rm s^{-1}$)& $11.58 \pm 0.06$ & $9.49 \pm 0.05$ & $7.63 \pm 0.05$\\
Trigger ($\rm s^{-1}$)& $12.22 \pm 0.06$ & $13.10 \pm 0.06$ & $8.20 \pm 0.05$\\
Trigger + ZS ($\rm s^{-1}$)& $13.01 \pm 0.06$ & $10.11 \pm 0.05$ & $8.79 \pm 0.05$\\
\hline
SN/Trigger& $0.948 \pm 0.007$ & $0.724 \pm 0.005$ & $0.930 \pm 0.008$\\
SN/(Trigger + ZS)& $0.890 \pm 0.006$ & $0.939 \pm 0.007$ & $0.868 \pm 0.008$\\
\hline
\end{tabular}
\end{table}

\begin{figure}[htbp]
\centering
        \subfigure[Y plane (standard trigger stream).]{
            \includegraphics[width=.47\linewidth]{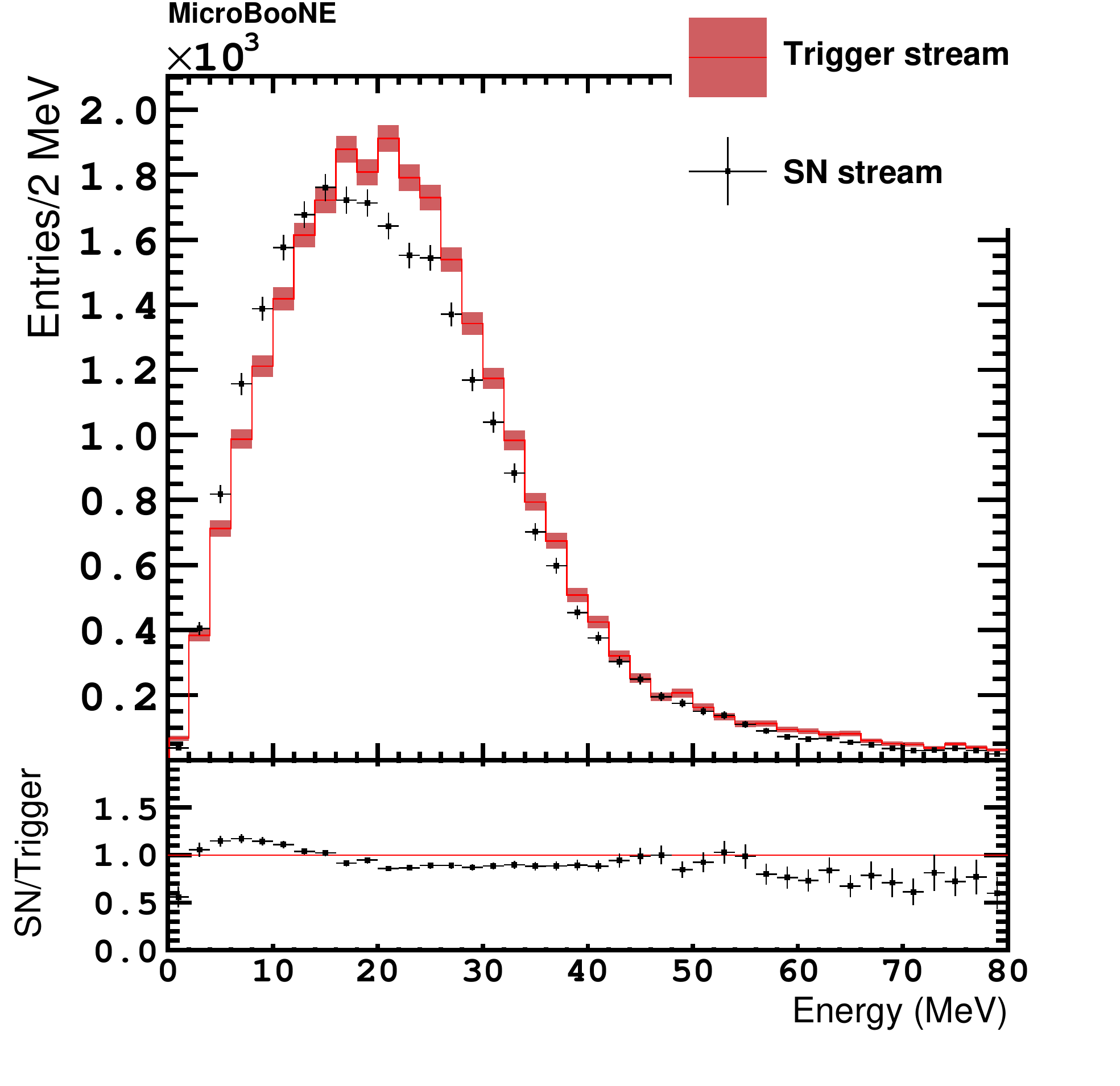}
            \label{fig:MichelIonizationSpectrum2}
        }
        \subfigure[Y plane (trigger stream with ZS emulation).]{
           \includegraphics[width=.47\linewidth]{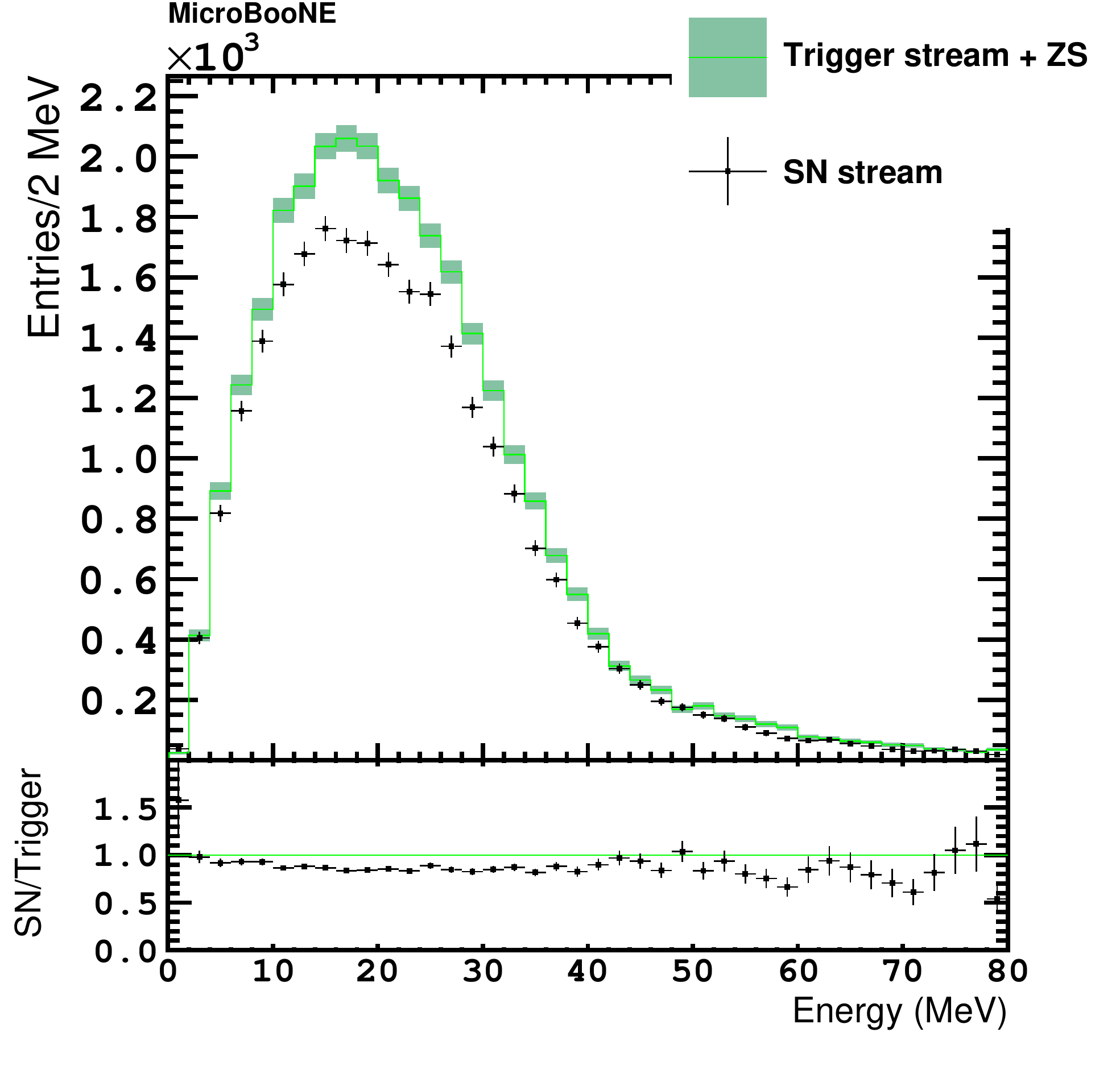}
            \label{fig:ZSMichelIonizationSpectrum2}
        }
	  \caption[Energy spectra of the Michel electron candidate ionization component on the collection plane.]{Energy spectra of the Michel electron candidate ionization component reconstructed using only the collection plane. The shift of the SN stream spectrum to lower energies in \subref{fig:MichelIonizationSpectrum2} is also seen on the induction planes and is well reproduced by the ZS emulation in \subref{fig:ZSMichelIonizationSpectrum2}. The markers and colors follow the same convention as figure~\ref{fig:MichelTotalSpectra2}.
	  }
	\label{fig:MichelIonizationSpectra2}
\end{figure}

\begin{figure}[htbp]
\centering
        \subfigure[Y plane (standard trigger stream).]{
            \includegraphics[width=.47\linewidth]{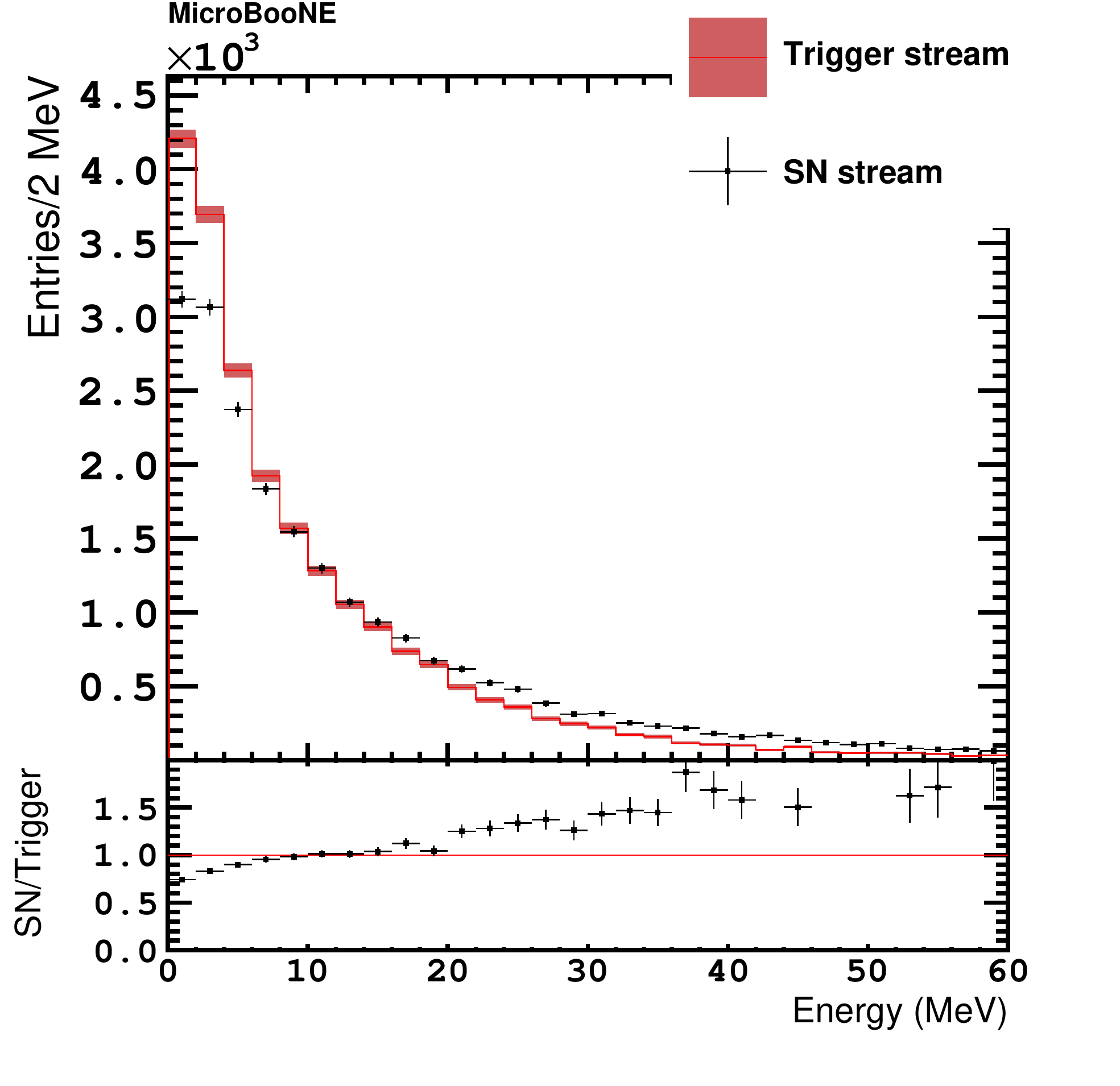}
            \label{fig:MichelRadiativeSpectrum2}
        }
        \subfigure[Y plane (trigger stream with ZS emulation).]{
           \includegraphics[width=.47\linewidth]{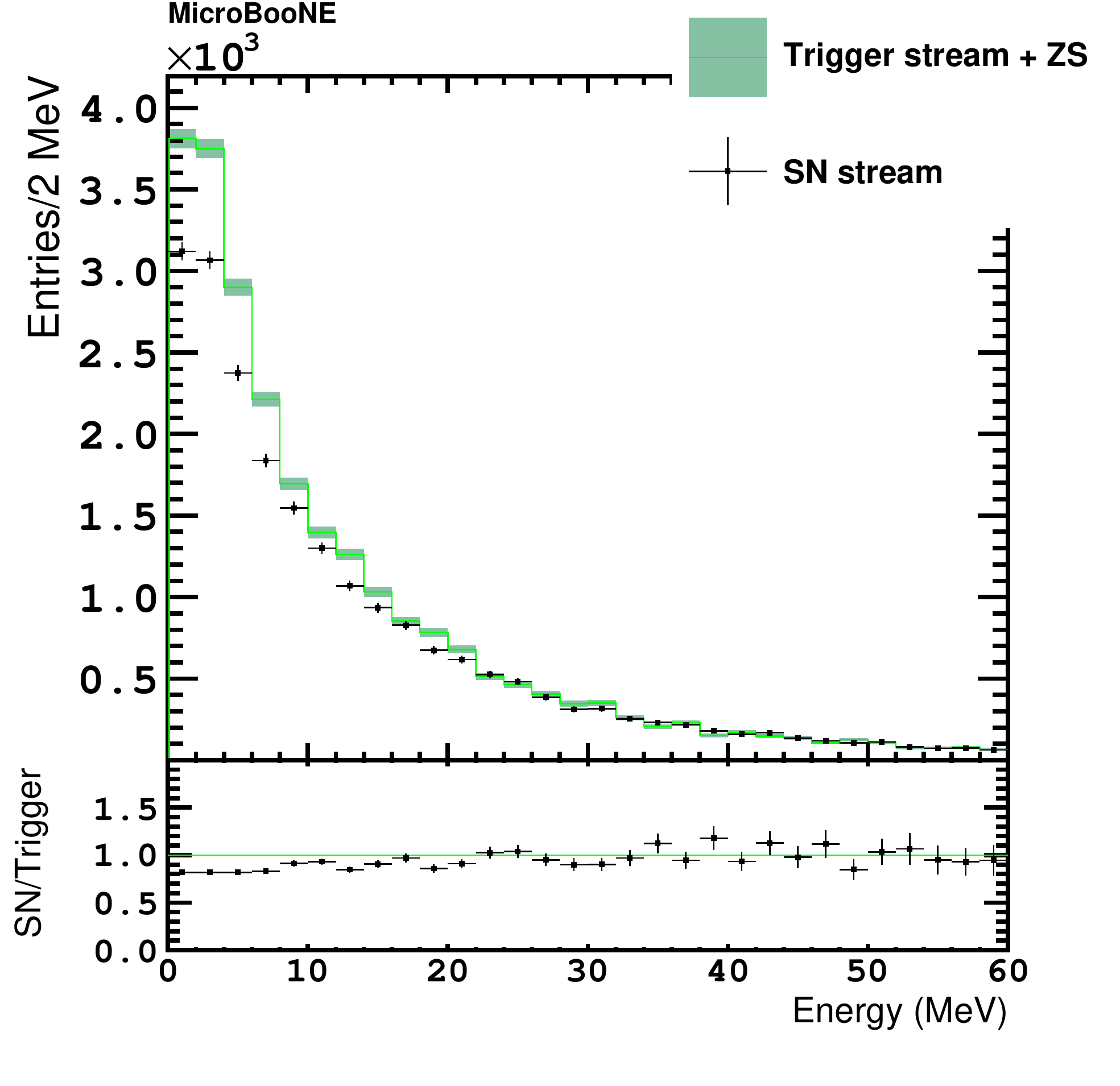}
            \label{fig:ZSMichelRadiativeSpectrum2}
        }
	  \caption[Energy spectra of the Michel electron candidate radiative component on the collection plane.]{Energy spectra of the Michel electron candidate radiative component reconstructed using only the collection plane. The SN stream spectrum shows an excess at high energy in \subref{fig:MichelRadiativeSpectrum2}, which is also present on the induction planes, and is well reproduced by the ZS emulation in \subref{fig:ZSMichelRadiativeSpectrum2}. The markers and colors follow the same convention as figure~\ref{fig:MichelTotalSpectra2}.
	  }
	\label{fig:MichelRadiativeSpectra2}
\end{figure}

\begin{figure}[htbp]
\centering
        \subfigure[Y plane (standard trigger stream).]{
            \includegraphics[width=.47\linewidth]{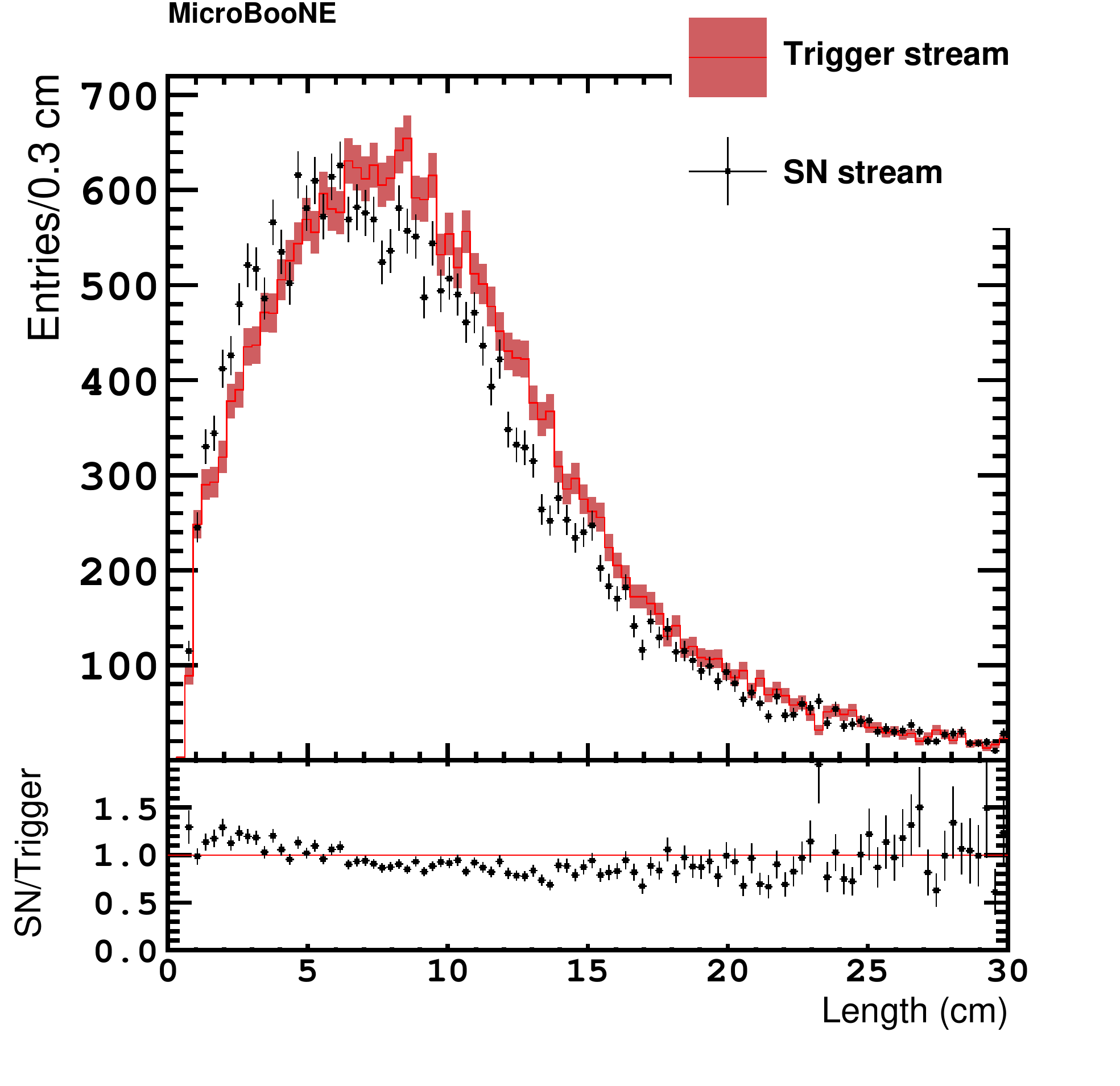}
            \label{fig:MichelLength2}
        }
        \subfigure[Y plane (trigger stream with ZS emulation).]{
           \includegraphics[width=.47\linewidth]{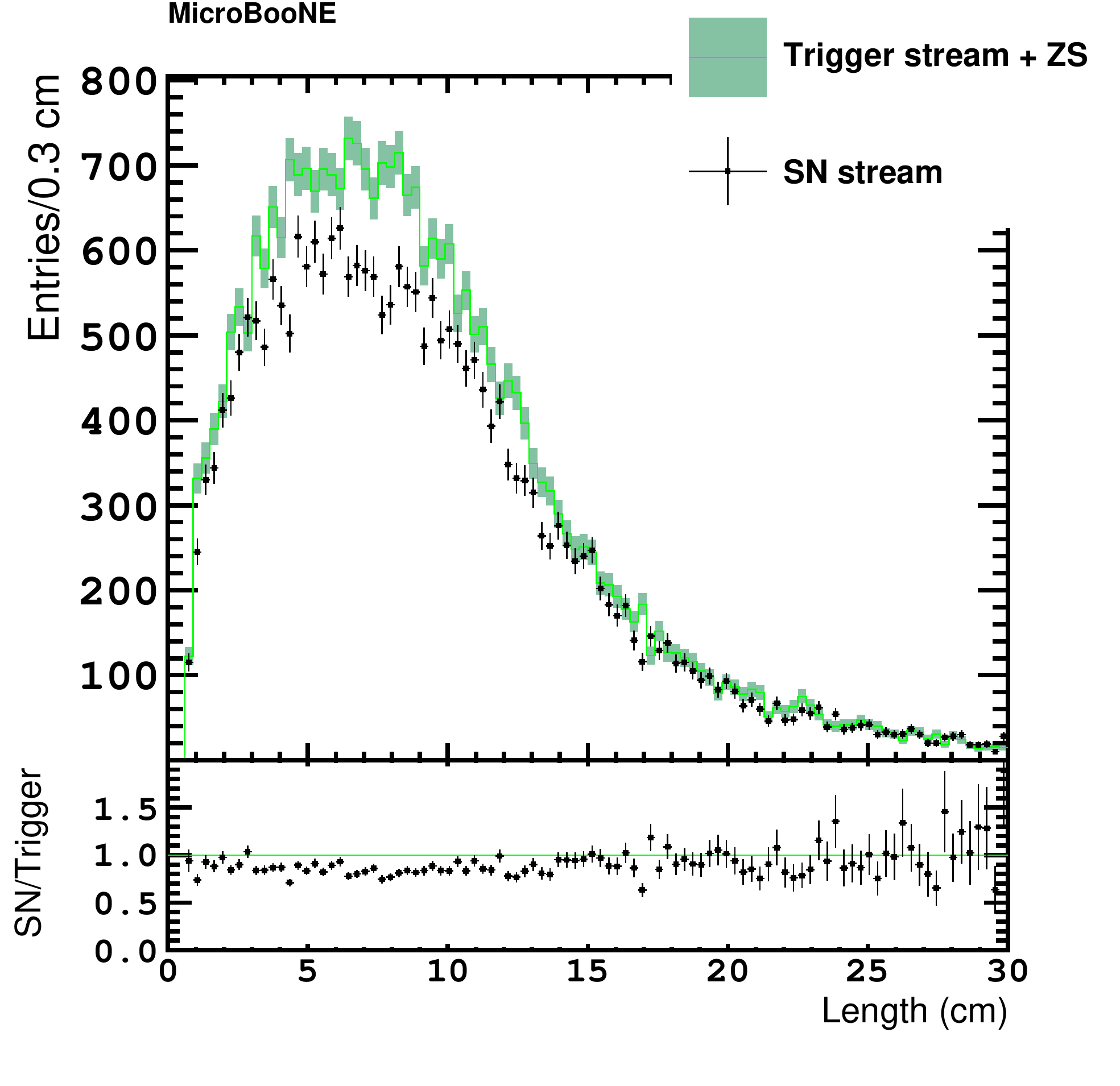}
            \label{fig:ZSMichelLength2}
        }
	  \caption[Length of the Michel electron candidate ionization component on the collection plane.]{Length of the Michel electron candidate ionization component on the collection plane. The shortened length of the ionization component of the SN stream candidates in \subref{fig:MichelLength2} is also seen on the induction planes, an effect which is reproduced by the ZS emulation in \subref{fig:ZSMichelLength2}. The markers and colors follow the same convention as figure~\ref{fig:MichelTotalSpectra2}.
	  }
	\label{fig:MichelLengthSpectra2}
\end{figure}

\begin{figure}[htbp]
\centering
        \subfigure[Y plane (standard trigger stream).]{
            \includegraphics[width=.47\linewidth]{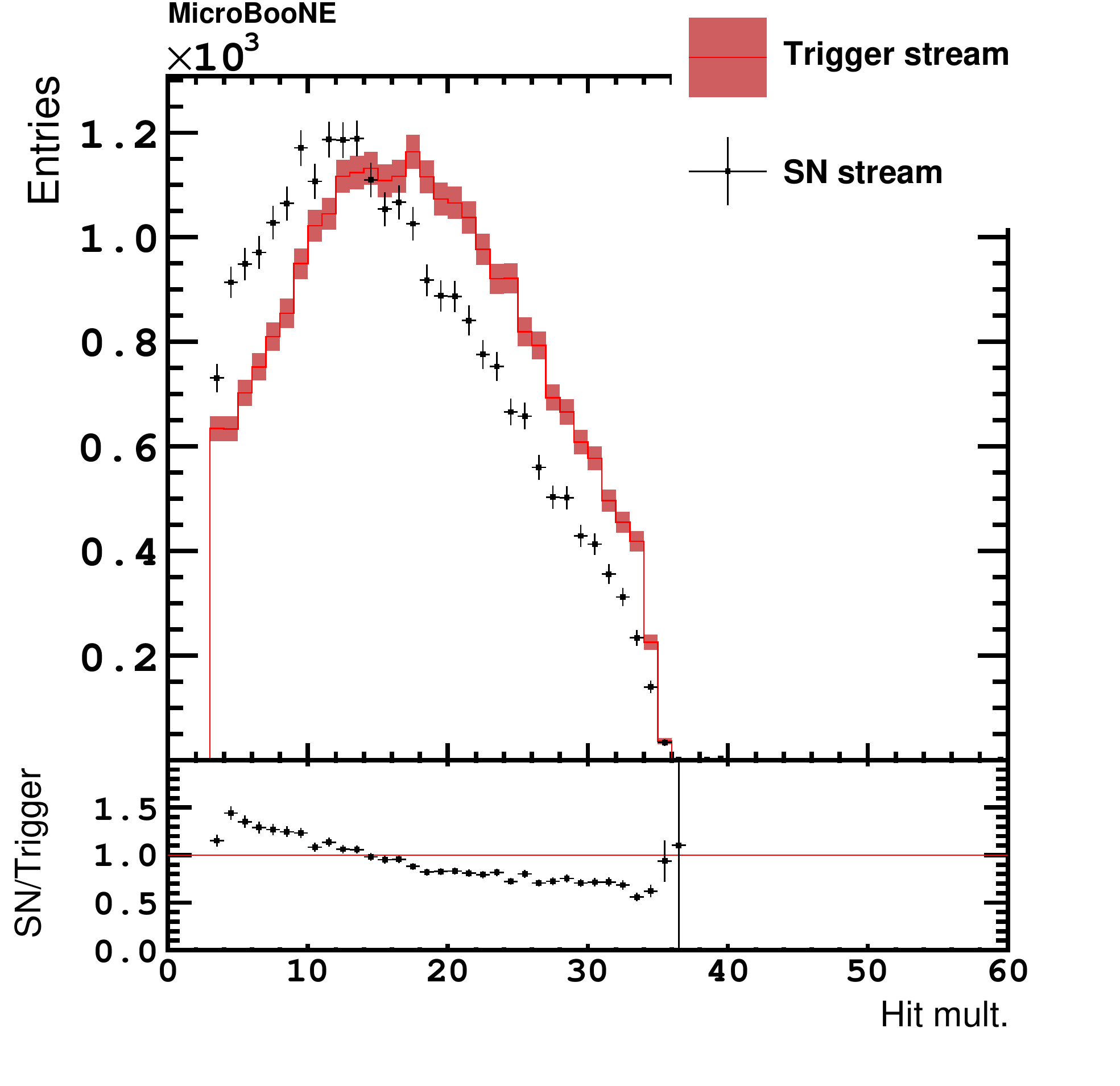}
            \label{fig:MichelIonizationHitMult2}
        }
        \subfigure[Y plane (trigger stream with ZS emulation).]{
           \includegraphics[width=.47\linewidth]{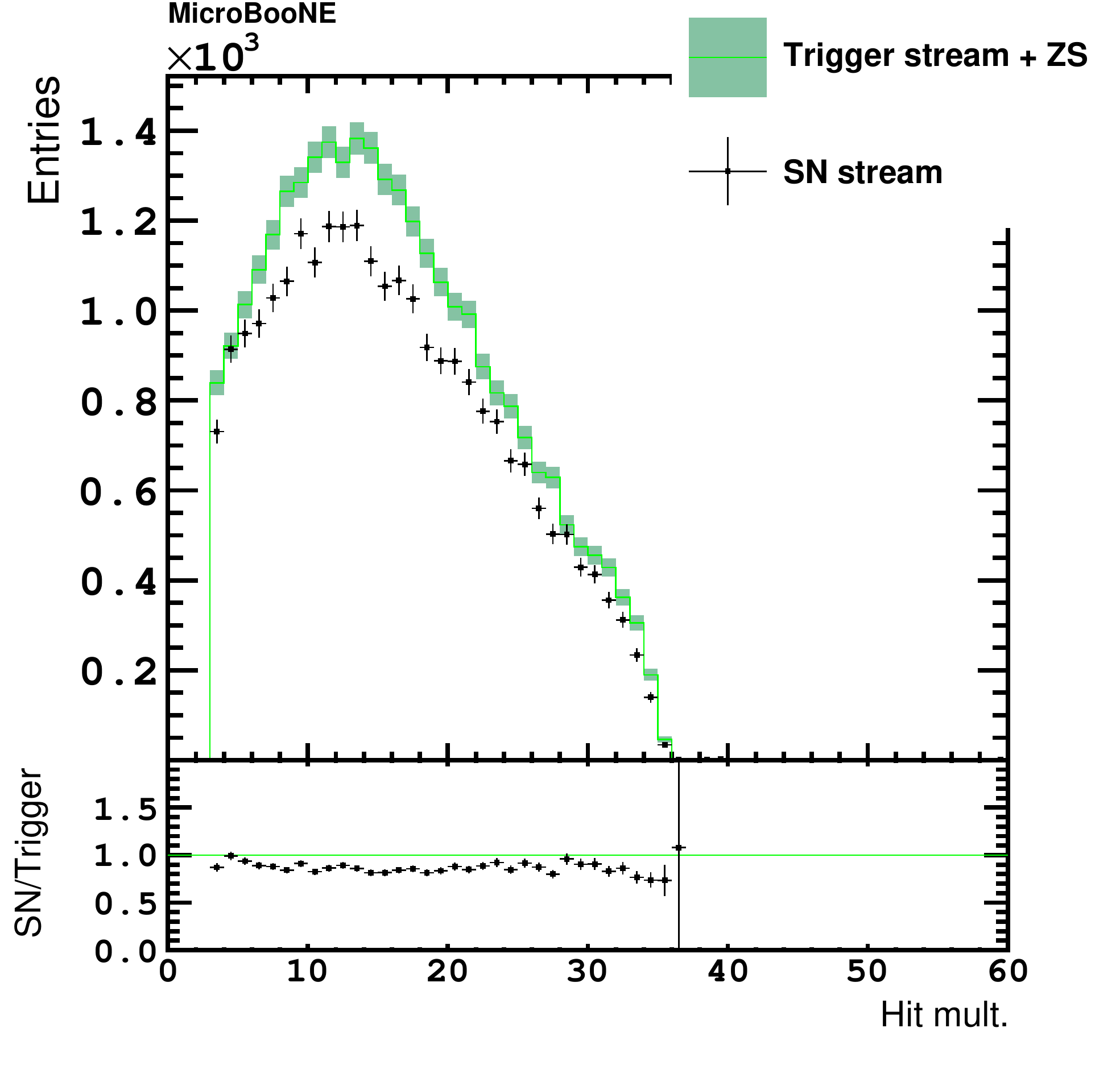}
            \label{fig:ZSMichelIonizationHitMult2}
        }
	  \caption[Hit multiplicity of the Michel electron candidate ionization component on the collection plane.]{Hit multiplicity of the Michel electron candidate ionization component reconstructed using only the collection plane. The shift to lower multiplicities observed in \subref{fig:MichelIonizationHitMult2} for the SN stream is well reproduced by the ZS emulation in \subref{fig:ZSMichelIonizationHitMult2}. The induction planes also show this shift in the SN stream. The markers and colors follow the same convention as figure~\ref{fig:MichelTotalSpectra2}.
	  }
	\label{fig:MichelIonizationHitMultCollection}
\end{figure}

\begin{figure}[htbp]
\centering
        \subfigure[Y plane (standard trigger stream).]{
            \includegraphics[width=.47\linewidth]{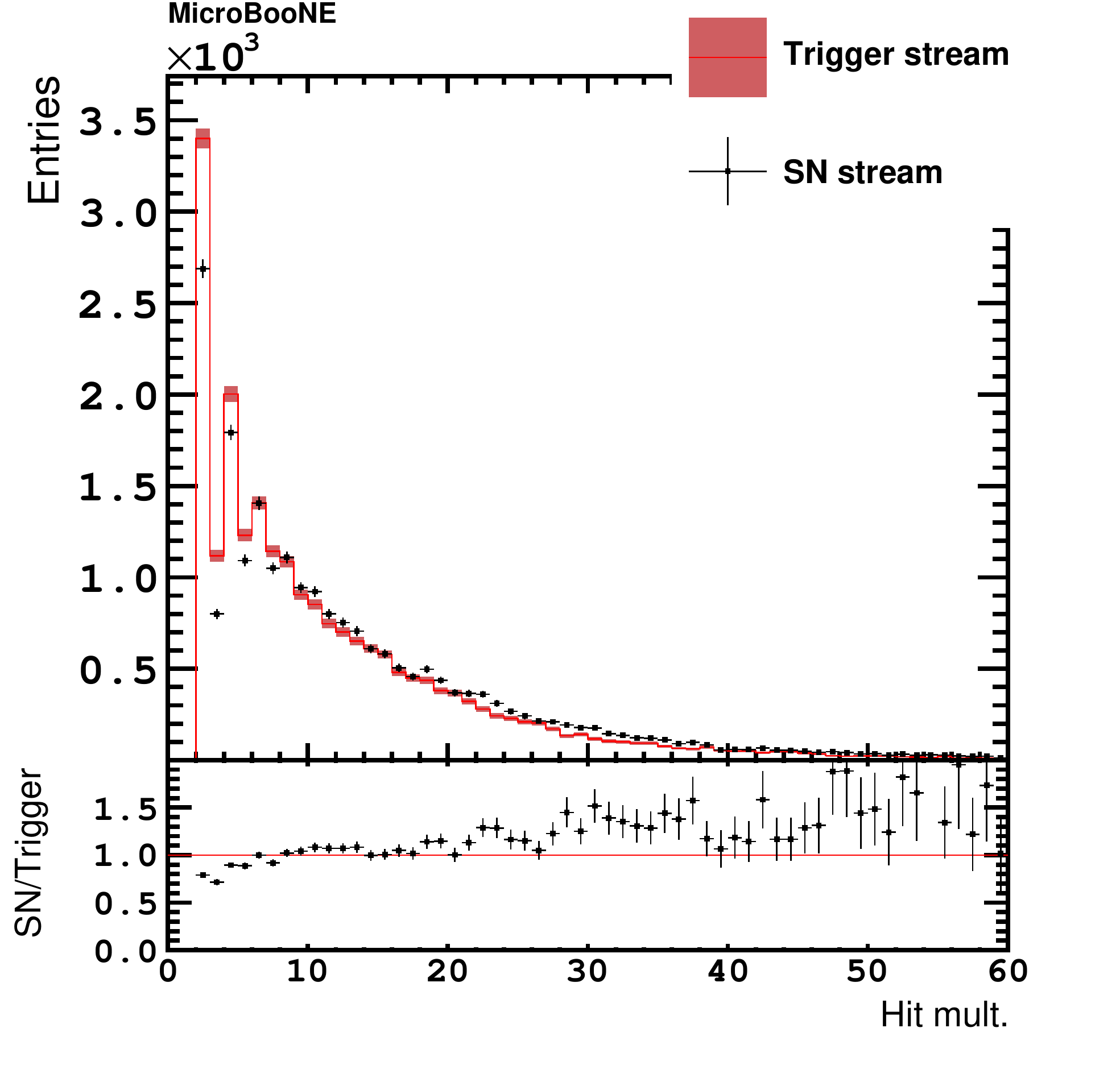}
            \label{fig:MichelRadiativeHitMult2}
        }
        \subfigure[Y plane (trigger stream with ZS emulation).]{
           \includegraphics[width=.47\linewidth]{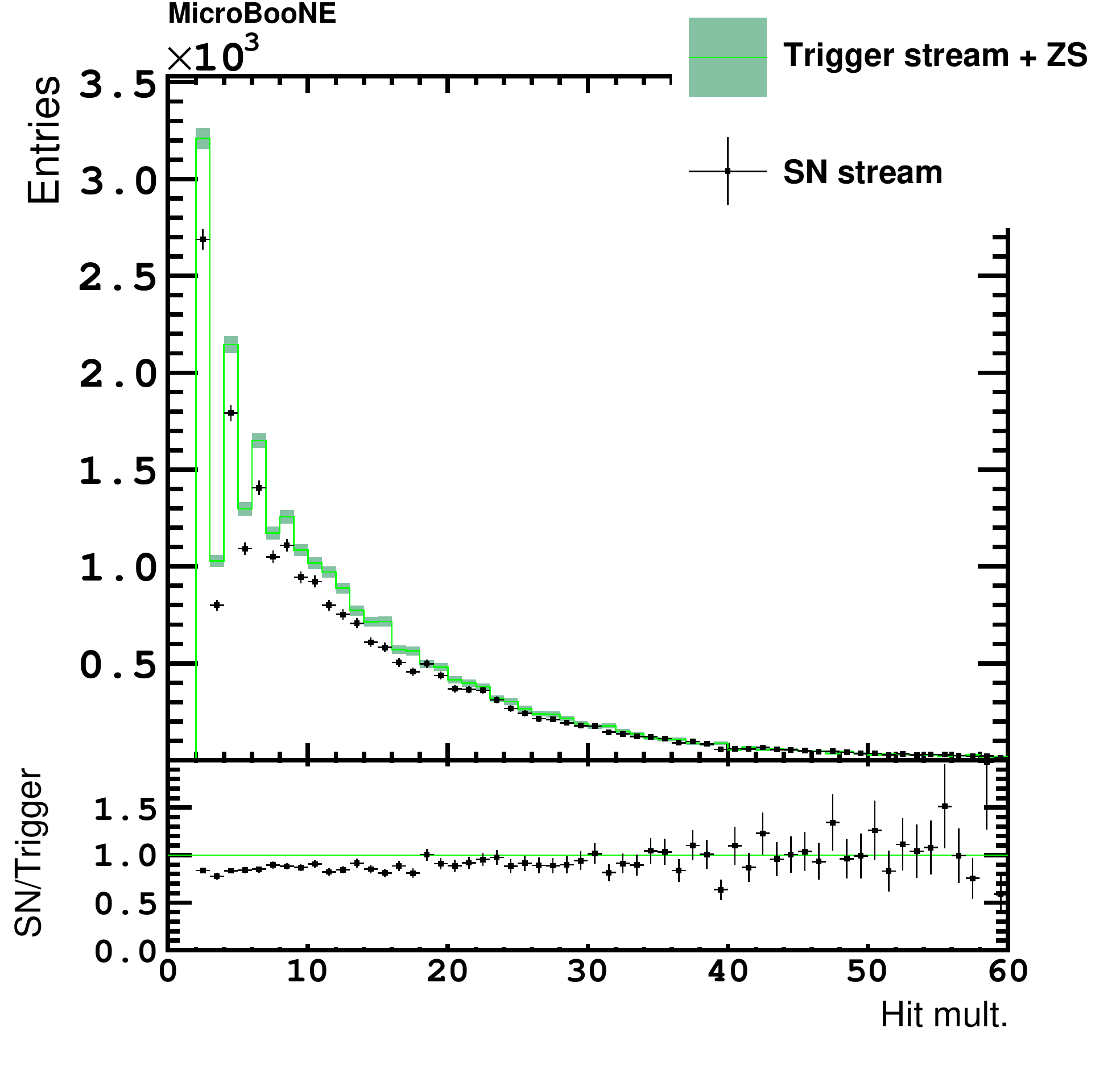}
            \label{fig:ZSMichelRadiativeHitMult2}
        }
	  \caption[Hit multiplicity of the Michel electron candidate radiative component on the collection plane.]{Hit multiplicity of the Michel electron candidate radiative component reconstructed using only the collection plane. \subref{fig:MichelRadiativeHitMult2} shows an excess at high multiplicity in the SN stream, which is also seen on the induction planes, and is well reproduced in \subref{fig:ZSMichelRadiativeHitMult2} by the ZS emulation. The markers and colors follow the same convention as figure~\ref{fig:MichelTotalSpectra2}.
	  }
	\label{fig:MichelRadiativeHitMultCollection}
\end{figure}

The effect of ZS is further studied by separating the ionization and radiative components of the Michel electrons. The ZS is found to cause a shift to lower energies of the ionization component (see figure~\ref{fig:MichelIonizationSpectra2}), and an excess of the radiative contribution at high energies (see figure~\ref{fig:MichelRadiativeSpectra2}). 
We interpret this effect as Michel electron tracks being split into segments by the ZS process. This leads to shorter reconstructed ionization components, confirmed explicitly by measuring the length of the ionization component shown in figure~\ref{fig:MichelLengthSpectra2}, and analyzing the hit multiplicity of the ionization component in figure~\ref{fig:MichelIonizationHitMultCollection}.
A consequence is the detached ionization segments being reconstructed as radiative components, increasing the radiative hit multiplicity as shown in figure~\ref{fig:MichelRadiativeHitMultCollection}.
These effects compensate each other when computing the total energy of the Michel electron by summing over all the ionization and radiative components. Overall, the zero suppression introduces a bias for the induction planes, $\approx 25\%$ for the U plane and $\approx 10\%$ for the V plane, but the bias for the collection plane is $< 1\%$. The resolution for the induction planes degrades by $\approx 20\%$, but the resolution for the collection plane, the most relevant for energy estimation, is not significantly affected.

The impact of ZS is stronger for the induction planes, as they feature smaller signals and higher thresholds (cf.~figure~\ref{fig:Thresholds18468}),
resulting in a relative shift of the distribution peaks to lower values when compared to the same distribution for the collection plane. Nevertheless, a similar shift is also seen in the trigger stream spectra (e.g.\ see figures~\ref{fig:MichelTotalSpectrum0} and~\ref{fig:MichelTotalSpectrum1} with respect to figure~\ref{fig:MichelTotalSpectrum2}).
This points to a higher inefficiency in collecting the charge on the induction planes which will also contribute to this shift.

\begin{figure}[htbp]
\centering
        \subfigure[U plane (standard trigger stream).]{
            \includegraphics[width=.47\linewidth]{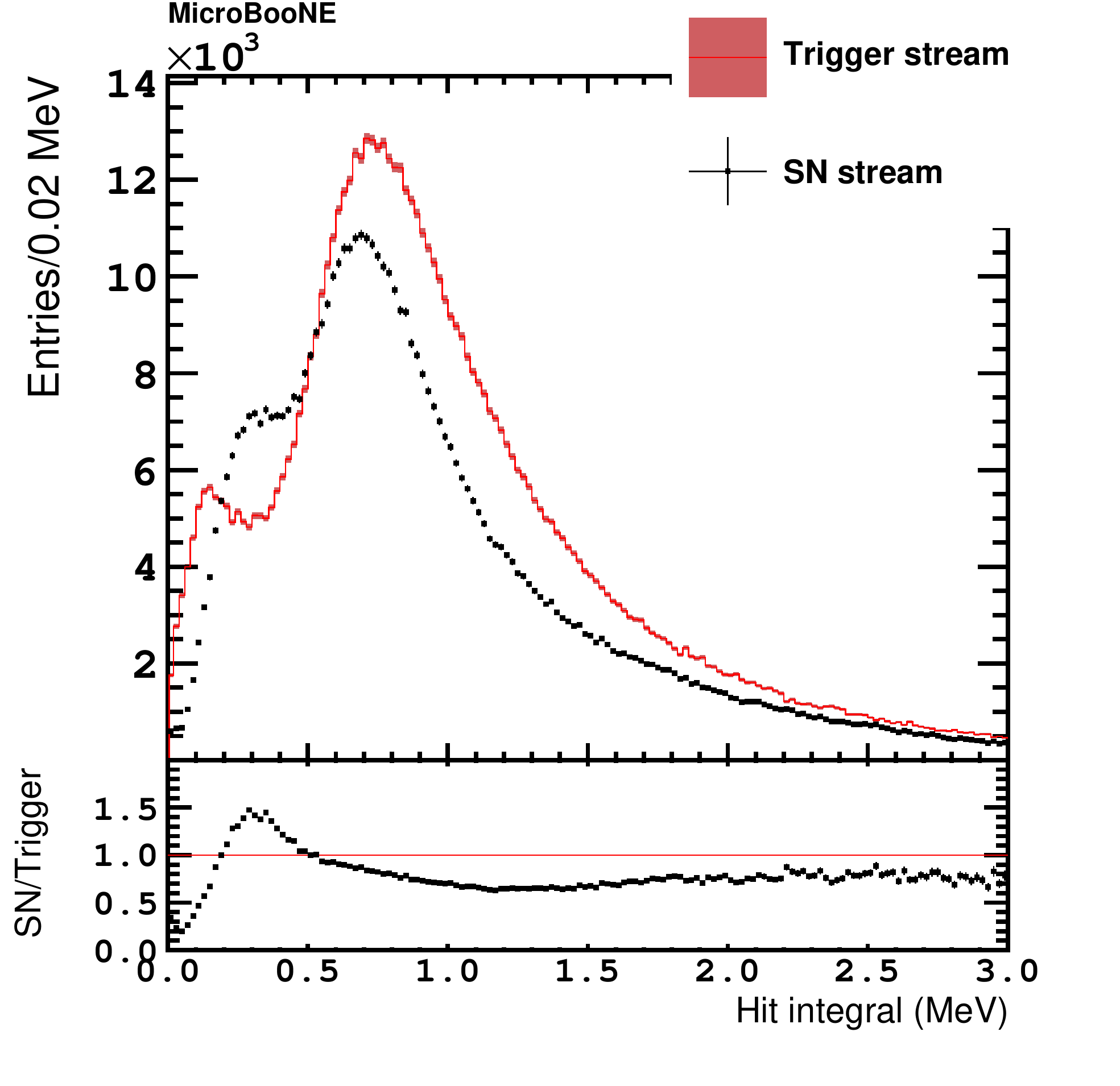}
            \label{fig:MichelIonizationHitSpectrum0}
        }
        \subfigure[U plane (trigger stream with ZS emulation).]{
           \includegraphics[width=.47\linewidth]{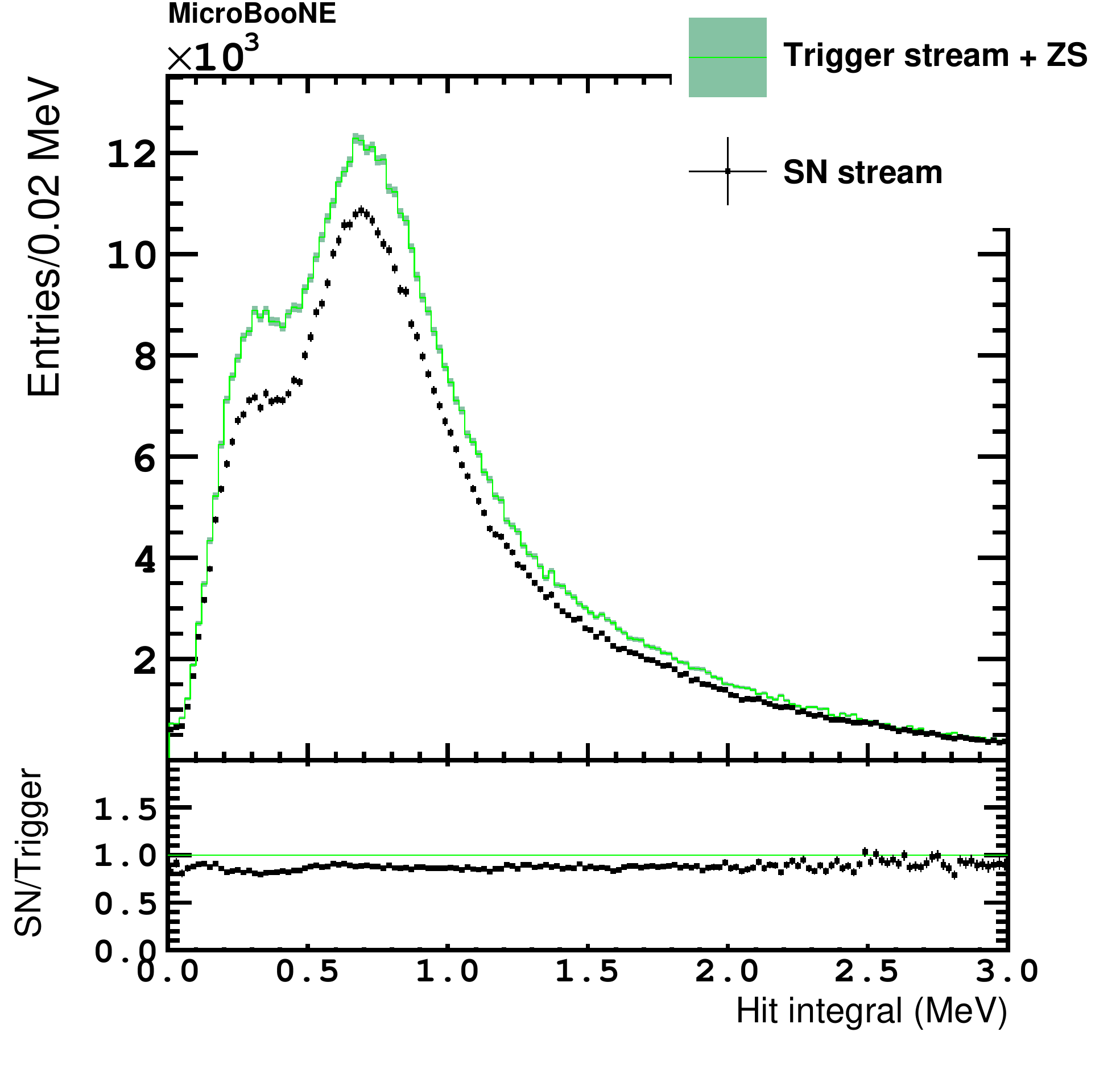}
            \label{fig:ZSMichelIonizationHitSpectrum0}
        }
        \subfigure[V plane (standard trigger stream).]{
            \includegraphics[width=.47\linewidth]{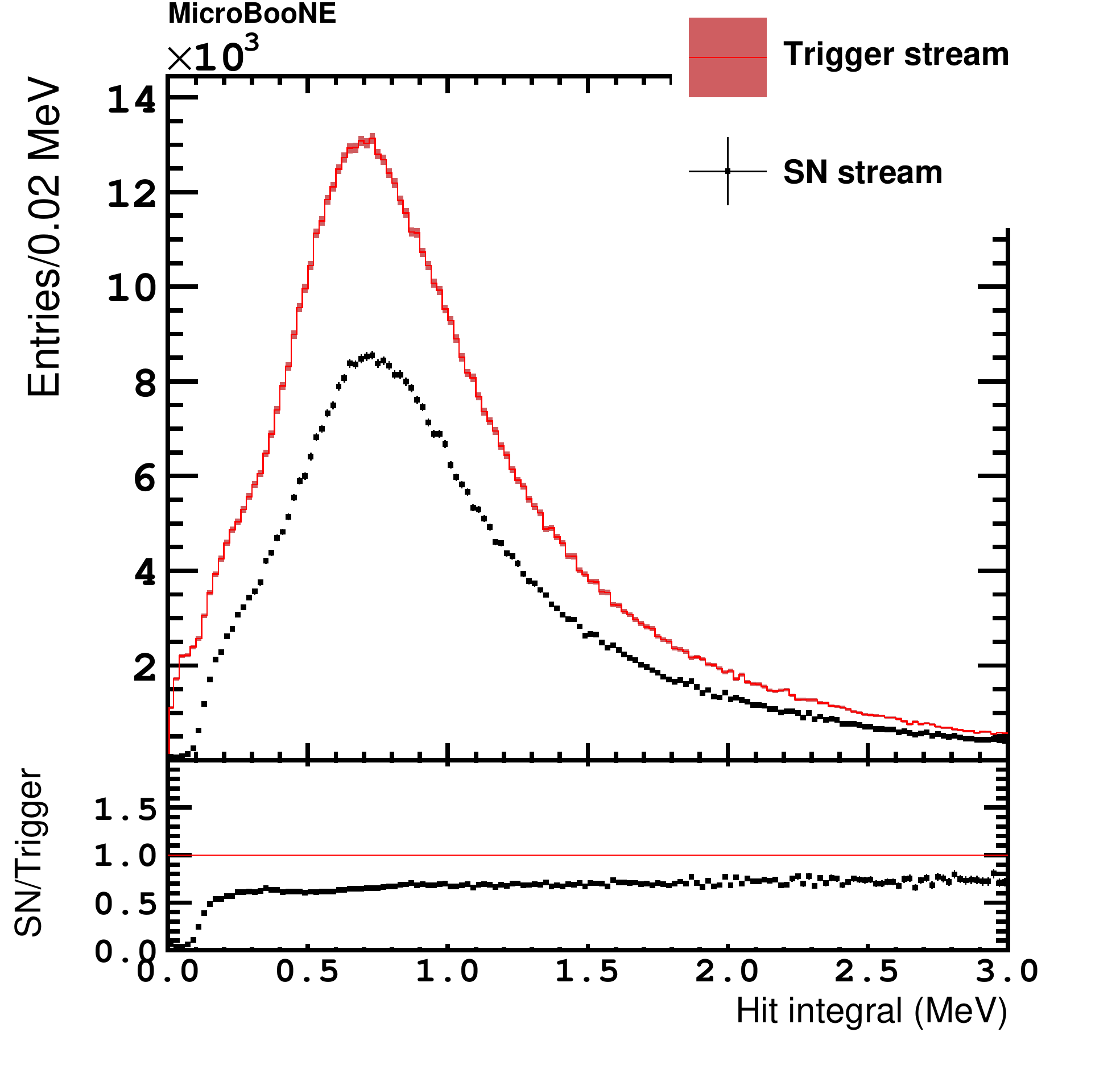}
            \label{fig:MichelIonizationHitSpectrum1}
        }
        \subfigure[V plane (trigger stream with ZS emulation).]{
           \includegraphics[width=.47\linewidth]{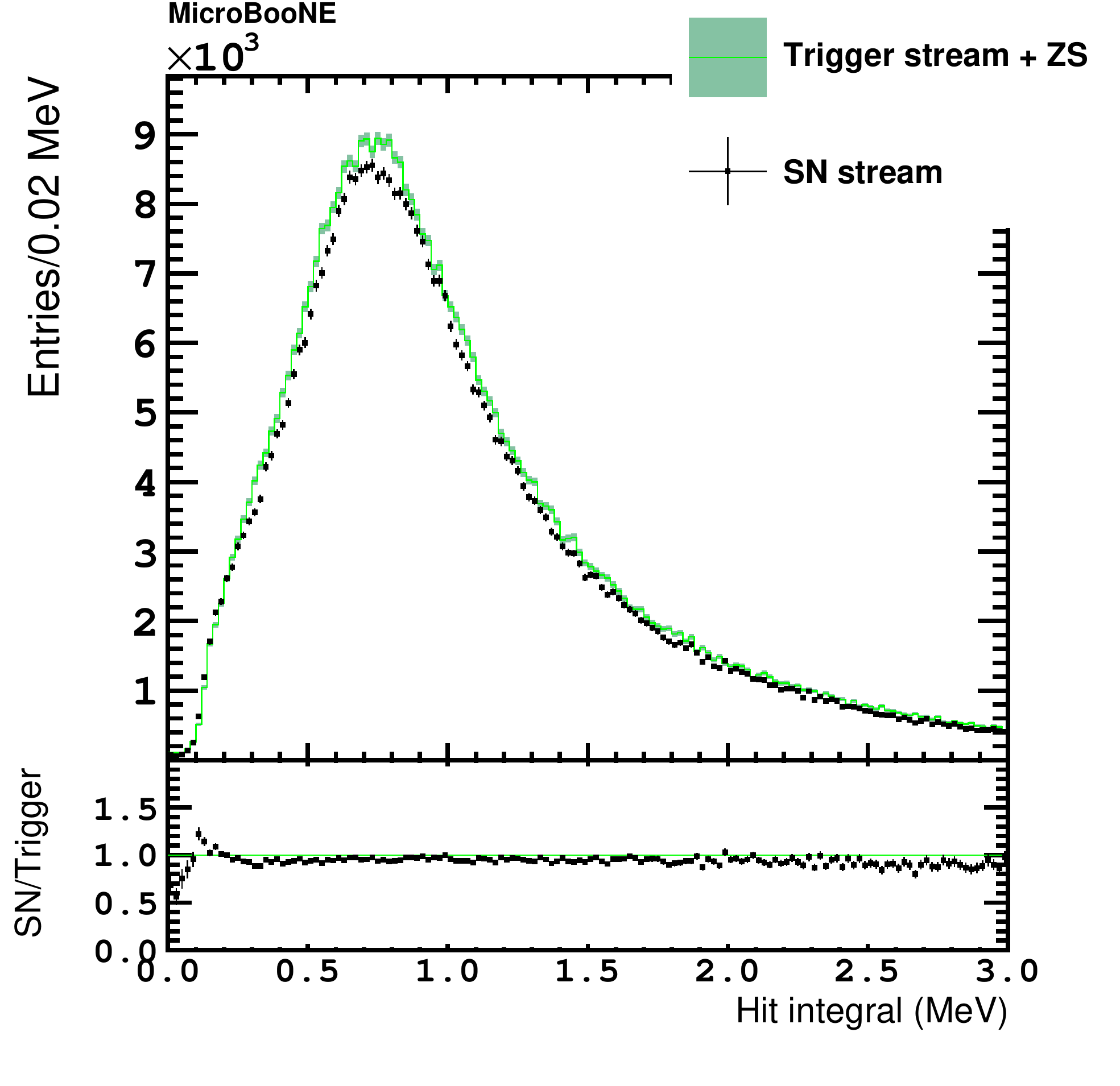}
            \label{fig:ZSMichelIonizationHitSpectrum1}
        }
	  \caption[Hit energy spectra of the Michel electron candidate ionization component on the induction planes.]{Hit energy spectra of the Michel electron candidate ionization component reconstructed using only one of the induction planes. The top row shows the first induction plane (plane U), the bottom row shows the second induction plane (plane V). For each row, the markers and colors follow the same convention as figure~\ref{fig:MichelTotalSpectra2}.
	  The SN stream data in \subref{fig:MichelIonizationHitSpectrum0} shows a low-energy tail above the trigger stream reference caused by the incomplete acquisition of the slow-rising pulses of the first induction plane resulting from the limited number of samples below threshold available in the FPGA implementation of ZS.
	  }
	\label{fig:MichelIonizationHitSpectraInduction}
\end{figure}

Among the induction planes, the U plane shows more extreme shifts to lower values due to ZS. This is expected since the U plane thresholds are higher than the V plane thresholds, but also because the charge from the slow-rising induction component of the U plane signals is not fully captured by the limited number of presamples allocated in the ZS.
In particular, the energy spectrum of the individual ionization hits, obtained from the ionization component of the Michel electron (see figure \ref{fig:evdMichelHitsBoxes_hits}), for the U plane (figure~\ref{fig:MichelIonizationHitSpectrum0}) shows an increase of the ``Compton-like'' tail with respect to the spectrum on the V plane (figure~\ref{fig:MichelIonizationHitSpectrum1}).

\begin{figure}[htbp]
\centering
        \subfigure[Y plane (standard trigger stream).]{
            \includegraphics[width=.47\linewidth]{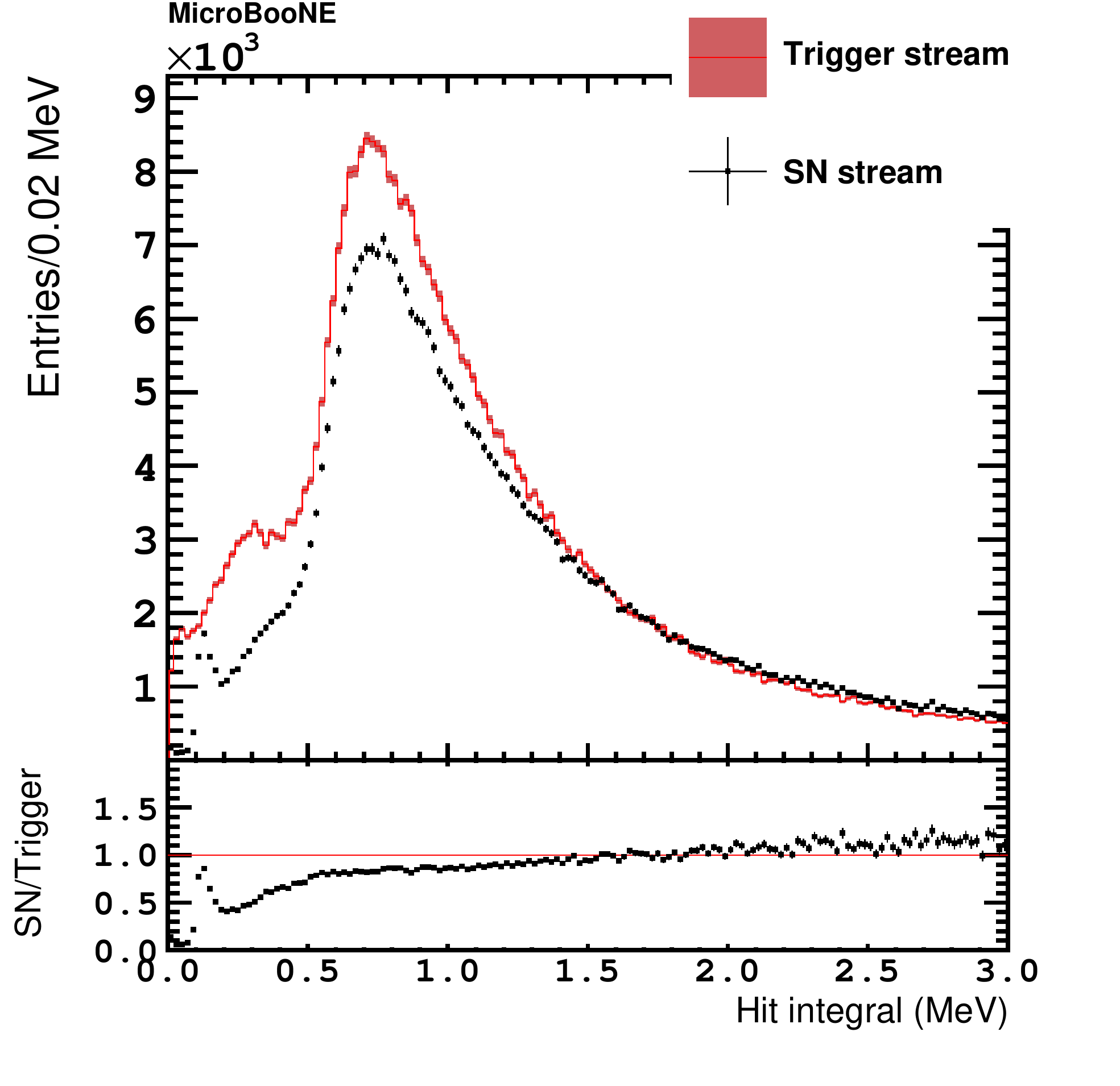}
            \label{fig:MichelIonizationHitSpectrum2}
        }
        \subfigure[Y plane (trigger stream with ZS emulation).]{
           \includegraphics[width=.47\linewidth]{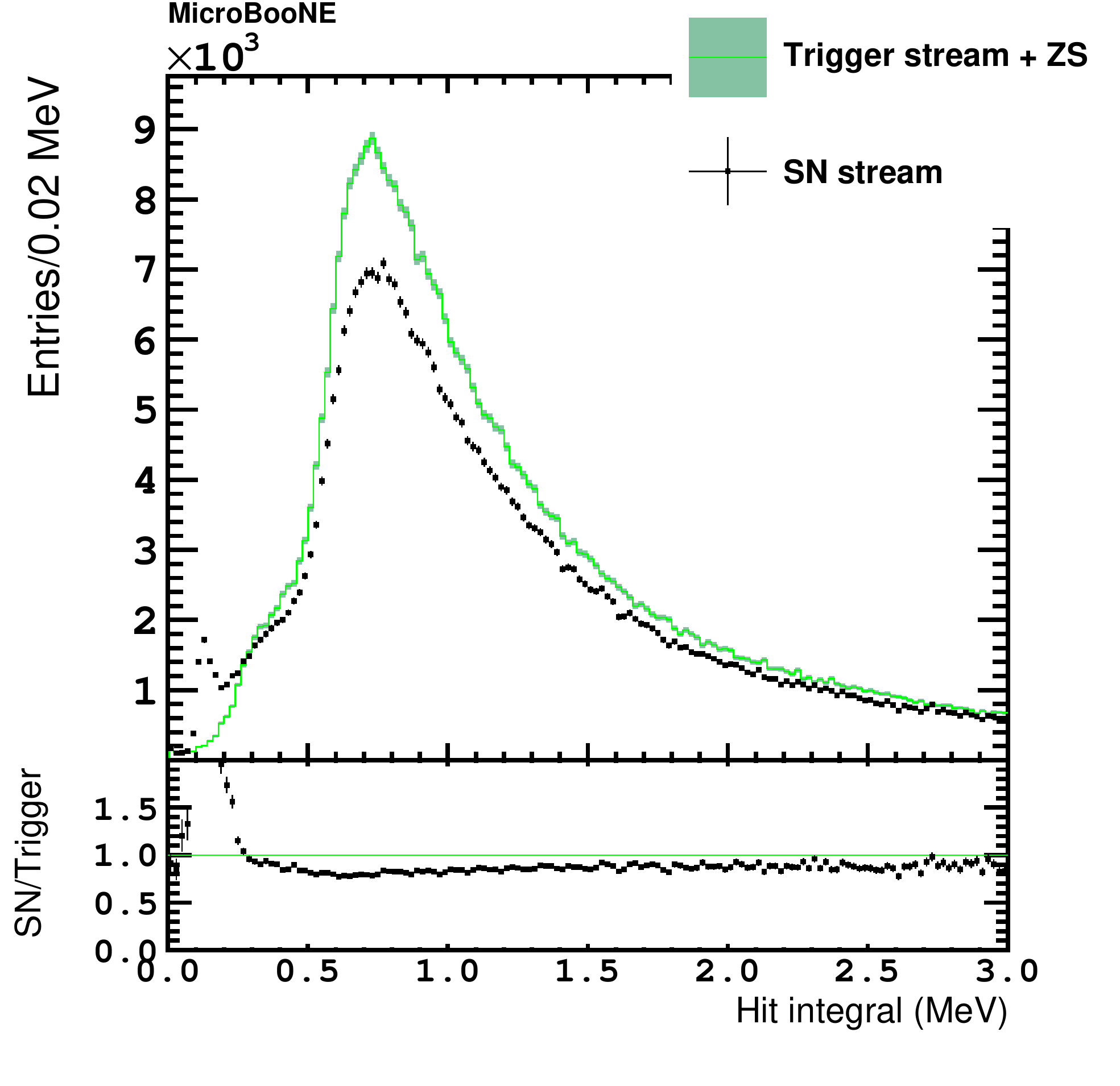}
            \label{fig:ZSMichelIonizationHitSpectrum2}
        }
	  \caption[Hit energy spectra of the Michel electron candidate ionization component on the collection plane.]{Hit energy spectra of the Michel electron candidate ionization component reconstructed using only the collection plane. The markers and colors follow the same convention as figure~\ref{fig:MichelTotalSpectra2}.
	  The peak at $0.1 - 0.2~\rm{MeV}$ found in the SN stream data is dominated by additional hits caused by flipped bits affecting the ADC words.
	  }
	\label{fig:MichelIonizationHitSpectra2}
\end{figure}

\begin{figure}[htbp]
\centering
        \subfigure[Y plane (standard trigger stream).]{
            \includegraphics[width=.47\linewidth]{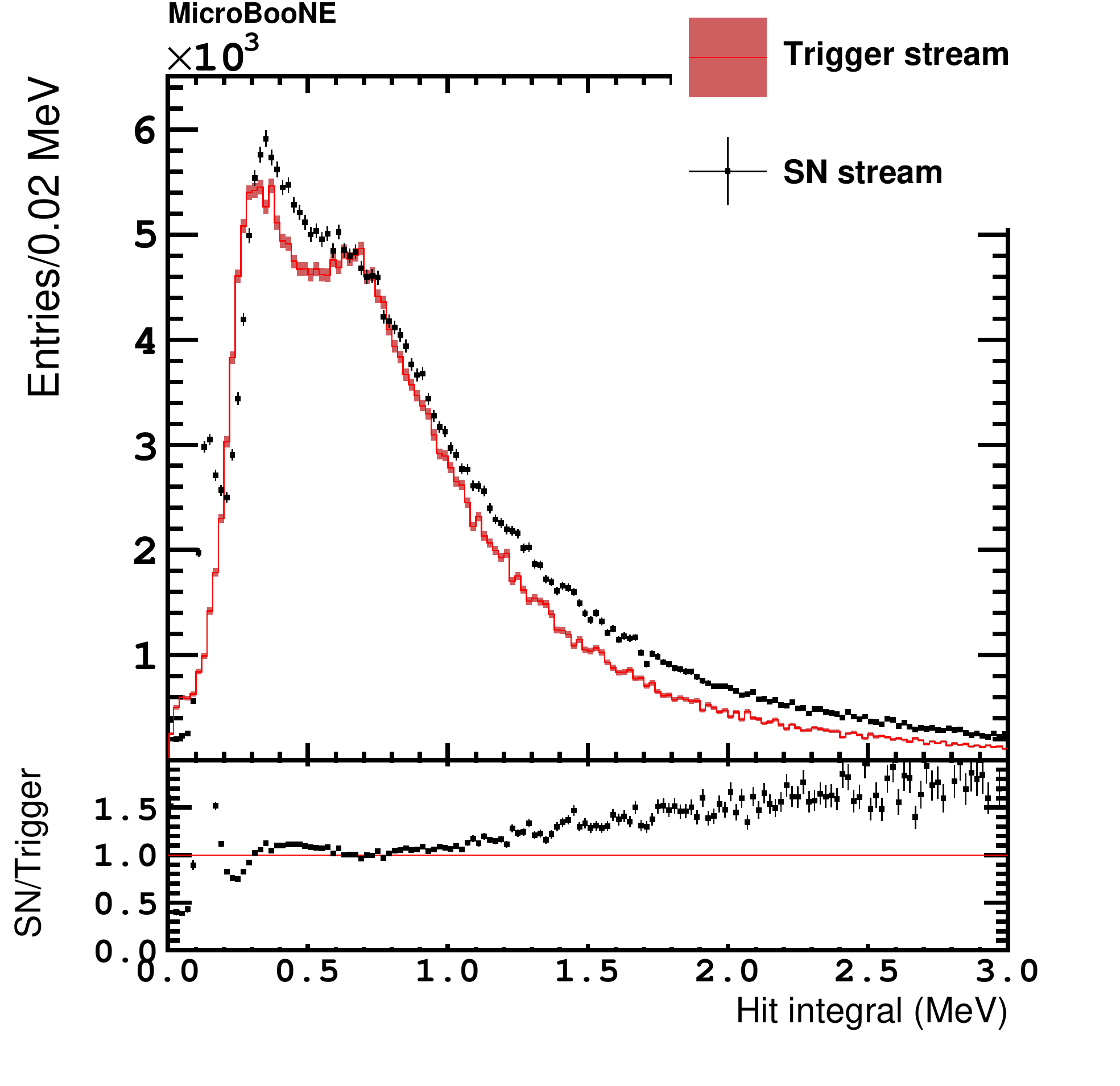}
            \label{fig:MichelRadiativeHitSpectrum2}
        }
        \subfigure[Y plane (trigger stream with ZS emulation).]{
           \includegraphics[width=.47\linewidth]{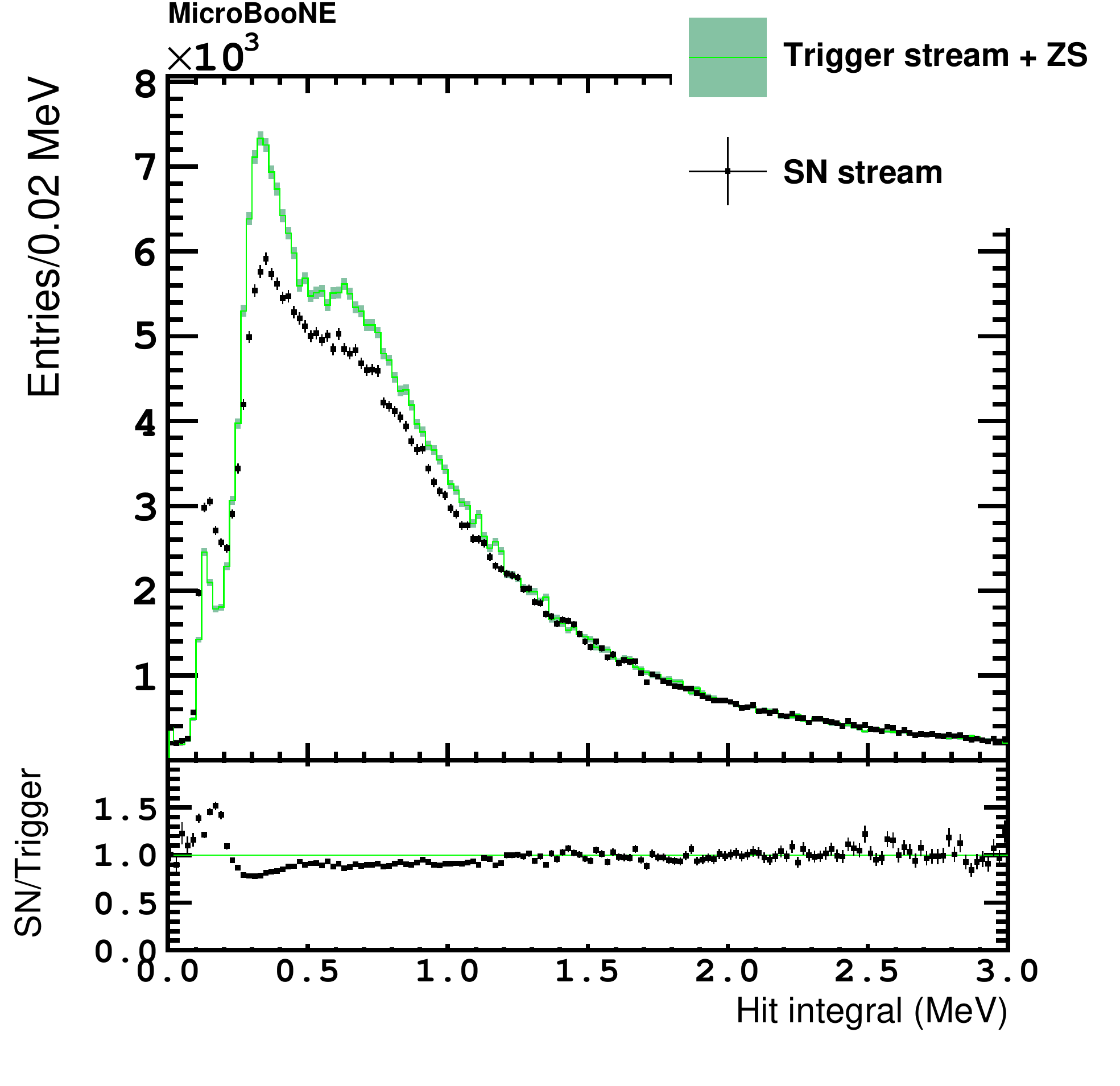}
            \label{fig:ZSMichelRadiativeHitSpectrum2}
        }
	  \caption[Hit energy spectra of the Michel electron candidate radiative component on the collection plane.]{Hit energy spectra of the Michel electron candidate radiative component reconstructed using only the collection plane. The markers and colors follow the same convention as figure~\ref{fig:MichelTotalSpectra2}.
	  The peak at $0.1 - 0.2~\rm{MeV}$ in the SN stream data is found to be caused by a combination of ZS and flipped bits affecting the ADC words. A similar effect is observed on the induction planes. 
	  }
	\label{fig:MichelRadiativeHitSpectra2}
\end{figure}

The good agreement shown by the trigger stream and simulated ZS with the SN stream hit-energy spectra allow us to anticipate that the impact of flipped bits on calorimetry after signal processing is very small. Figures~\ref{fig:MichelIonizationHitSpectra2},~\ref{fig:MichelIonizationHitSpectraInduction}, and~\ref{fig:MichelRadiativeHitSpectra2} show the effect of flipped bits as a small peak at $0.1 - 0.2~\rm{MeV}$, which is more prominent on the collection plane, where flipped bits have been found to be $\approx 9\%$ more frequent.
This is understood to be caused by the flipped bits which escape correction by the filter and distort the waveform, forcing the hit finder to allocate extra hits to fit the waveform shape. Because only small shifts in ADC counts escape correction, these additional hits have small amplitudes.
Using a data-driven simulation of the flipped bits we have evaluated the impact on the hit energy resolution to be $\approx 10\%$. Note these additional hits are effectively rejoined when summing over all the hits within the cluster to estimate its energy, resulting in a similar impact on the event. Due to the dominant ($\approx 20\%$) contribution to the resolution caused by the failure to reconstruct very low energy photons radiated by the electrons~\cite{Acciarri:2017sjy}, we deem this additional contribution acceptable, even though the investigation of the origin of the flipped bits continues. In addition, not all the hits contributing to the $0.1 - 0.2~\rm{MeV}$ peak come from flipped bits, as the ZS is found to also create additional radiative-like hits in that region, shown by the peak found in the trigger stream distributions with simulated ZS in figure~\ref{fig:ZSMichelRadiativeHitSpectrum2}, for which flipped bits were not simulated.

\section{The Continuous Readout Stream as a development platform}
\label{sec:DevelopmentPlatform}
The MicroBooNE SN stream is the first realized stage toward a self-triggering LArTPC based on TPC information. Its successful operation enables both physics measurements and a platform to develop and test TPC-based data selection algorithms for current and future detectors. For example, the zero-suppressed ROIs can be processed offline to extract features to be used as trigger primitives, which then would be clustered and processed through pattern recognition. See reference~\cite{REU2019} for an example using the DUNE trigger primitive prescription~\cite{Abi:2020loh}. A detailed discussion is beyond the scope of this work, but our analysis of Michel electrons already suggests that the energy bias observed in the reconstruction of the ionization component is an effect to be taken into account when defining the thresholds for such a trigger. 
The possibility of triggering using any of the TPC planes, especially for enabling an online 3-D hit reconstruction based on matching time and wire coordinates, which can reduce the impact of noise and ambiguities when clustering, is very attractive. In this regard, we observe a large loss ($\approx 20\%$) of reconstructed Michel electrons on the second induction plane caused by the ZS, which would decrease the trigger efficiency for this plane. 
The second induction plane is shielded by the first induction plane, resulting in signals with smaller amplitudes, which are more difficult to separate from the electronics noise, and symmetrical, which makes them prone to cancellation due to destructive interference.
While the long induction rising edge of the signals on the first induction plane may be challenging to capture, the asymmetrical nature of the pulses on this plane 
makes them more suitable for triggering. The inefficiency of a shielded plane can be partially mitigated if a peak-signal-to-noise ratio large enough is achieved (for reference, MicroBooNE U and V planes have $S/N$ of $18.1$ and $13.1$, respectively~\cite{Acciarri:2017sde}), which would allow the setting of lower ZS thresholds.

\section{Conclusion}
MicroBooNE is the first liquid argon time projection chamber experiment to successfully commission and operate a continuous readout, opening a new way to look at data from the MicroBooNE detector.
This novel data stream gives MicroBooNE the possibility to detect the burst of core-collapse supernova neutrinos using the SNEWS alert as a delayed trigger, expanding the physics program of the experiment. We defer the discussion of reconstruction and selection of those neutrinos to future work.
After one and a half years of successful operation, during which we have tested multiple FPGA-based compression algorithms, the goal of reaching a stable compression factor $\approx 80$, with sensitivity to supernova neutrino energies, has been accomplished on all the TPC planes, including the induction planes where the pulse shapes are more challenging.
The best performance was found for a zero-suppression algorithm that employs static baselines and bandwidth-driven individual thresholds for each channel. 

Based on the rate of Michel electrons from stopping cosmic-ray muons reconstructed in the continuous readout stream, relative to the rate observed in the trigger stream, we estimate a relative detection efficiency of $(93.0 \pm 0.8)\%$ on the collection plane, $(72.4 \pm 0.5)\%$ on the second induction plane, and $(94.8 \pm 0.7)\%$ on the first induction plane, where the uncertainties are statistical, and do not include systematic effects such as 
seasonal variations in the cosmic ray rate.

An unexpected challenge in the implementation of the continuous readout stream is the appearance of flipped bits, affecting $\approx 4\%$ of the ADC samples. Their origin is still being investigated, but their effect has resulted in an acceptable contribution to the energy resolution ($\approx 10\%$) due to several mitigation steps in the offline reconstruction. Efforts continue to understand and correct for this effect.

The continuous readout of a LArTPC is the first stage towards eventually developing a trigger based on the ionization patterns observed in the TPC. Analysis of the Michel electrons reveals that the zero suppression impacts the reconstruction of the ionization component of the electrons, resulting in reconstructed lower energies. This effect is well reproduced by our readout simulation. While this energy bias would have little impact on the measurement of the energy of supernova neutrinos, where the total energy is computed by including radiative-like components, it is an effect that would have to be accounted for when setting thresholds for an hypothetical TPC-based trigger that uses the zero-suppressed ionization clusters. We also observe a large inefficiency ($\approx 20\%$) in the reconstruction of Michel electrons on the second induction plane caused by the combination of the smaller signals due to screening by the first induction plane and the zero suppression. 

Whereas this work focuses on using the SN stream to detect low-energy electrons such as those produced by supernova neutrinos, 
MicroBooNE can use the SN stream data to make novel measurements of other off-beam physics such as nucleon decay (proton decay, neutron-antineutron oscillation, etc) in a liquid argon TPC and is a platform to develop analyses for future detectors and to study possible backgrounds~\cite{Hewes:2017xtr}.

%\clearpage
\appendix

\section{Dynamic-baseline algorithm}
\label{app:DynamicBaseline}
\begin{figure}[htbp]
\centering
\includegraphics[width=0.7\textwidth]{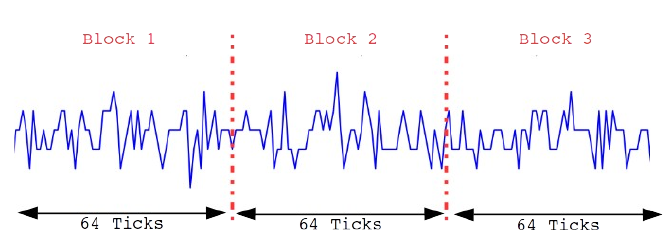} 
\caption[Dynamic baseline estimation.]{Cartoon showing how the waveform is split into blocks for the dynamic baseline estimation.}
\label{fig:ZeroSuppressionExampleBaseline}
\end{figure}

This appendix describes the algorithm implemented in the FEM FPGA to estimate the baseline in real time. The estimation of the channel baseline is performed using 3~contiguous blocks of 64~samples each (each block corresponding to $32~\mu\rm{s}$ of the waveform, see figure~\ref{fig:ZeroSuppressionExampleBaseline}). 
A rounded mean ADC value for each block is computed by summing the ADC values of all 64~samples and then dropping the 6~least significant bits (equivalent to an integer division by $64$). 
A truncated ADC variance for each block is computed by summing the squared differences between the ADC value of each sample and the rounded mean computed above, and then dropping the 6~least significant bits. 
If the absolute value of the difference between an ADC sample and the rounded mean is greater than or equal to 63~ADC counts, the contribution of that sample to the variance is fixed to 4095~ADC counts to prevent arithmetic overflows. 
The rounded mean and the truncated variance of each block are compared to the values from the other two blocks to avoid choosing a baseline corresponding to an actual signal or a non-representative fluctuation. 
If the three rounded mean differences and the three truncated variance differences between blocks are within the configurable tolerance values, the rounded mean of the central block is taken as the new baseline and applied for ZS for ADC samples beginning after the third block. 
The local baseline estimation is applied continuously as a sliding window from the beginning of the run, dropping the oldest 64-sample block and adding a newer block. 
The baseline tolerance parameters (rounded mean and truncated variance differences) are configured per FEM (i.e. in groups of 64~channels). 
If the baseline conditions are not satisfied (e.g.\ the difference between blocks never meets the tolerance) in a channel, the ZS algorithm does not produce output data for that channel.

\section{Physics-driven plane-wide zero-suppression thresholds}
\label{app:PlanewiseThresholds}
This appendix describes the first method used to determine the settings for the ZS of TPC waveforms. This method aims at establishing a single threshold for each TPC plane to separate signals (dominated by cosmic-ray muons) from electronics noise.

\begin{figure}[htbp]
\centering
\subfigure[Maximum ADC values.]{
  \includegraphics[width=0.7\textwidth]{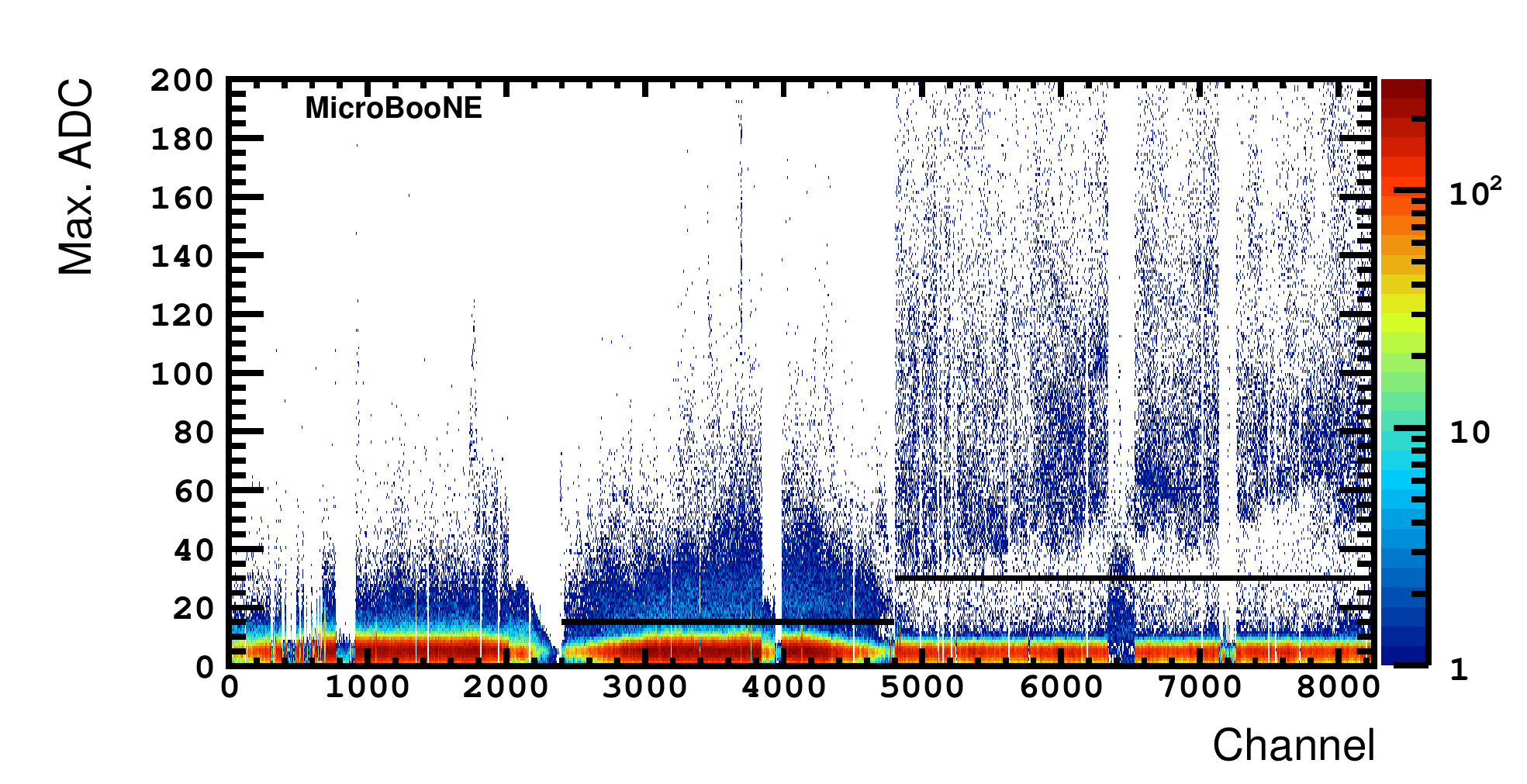}
  \label{fig:ADCSpectraPerPlane_max}
}
\subfigure[Minimum ADC values.]{
  \includegraphics[width=0.7\textwidth]{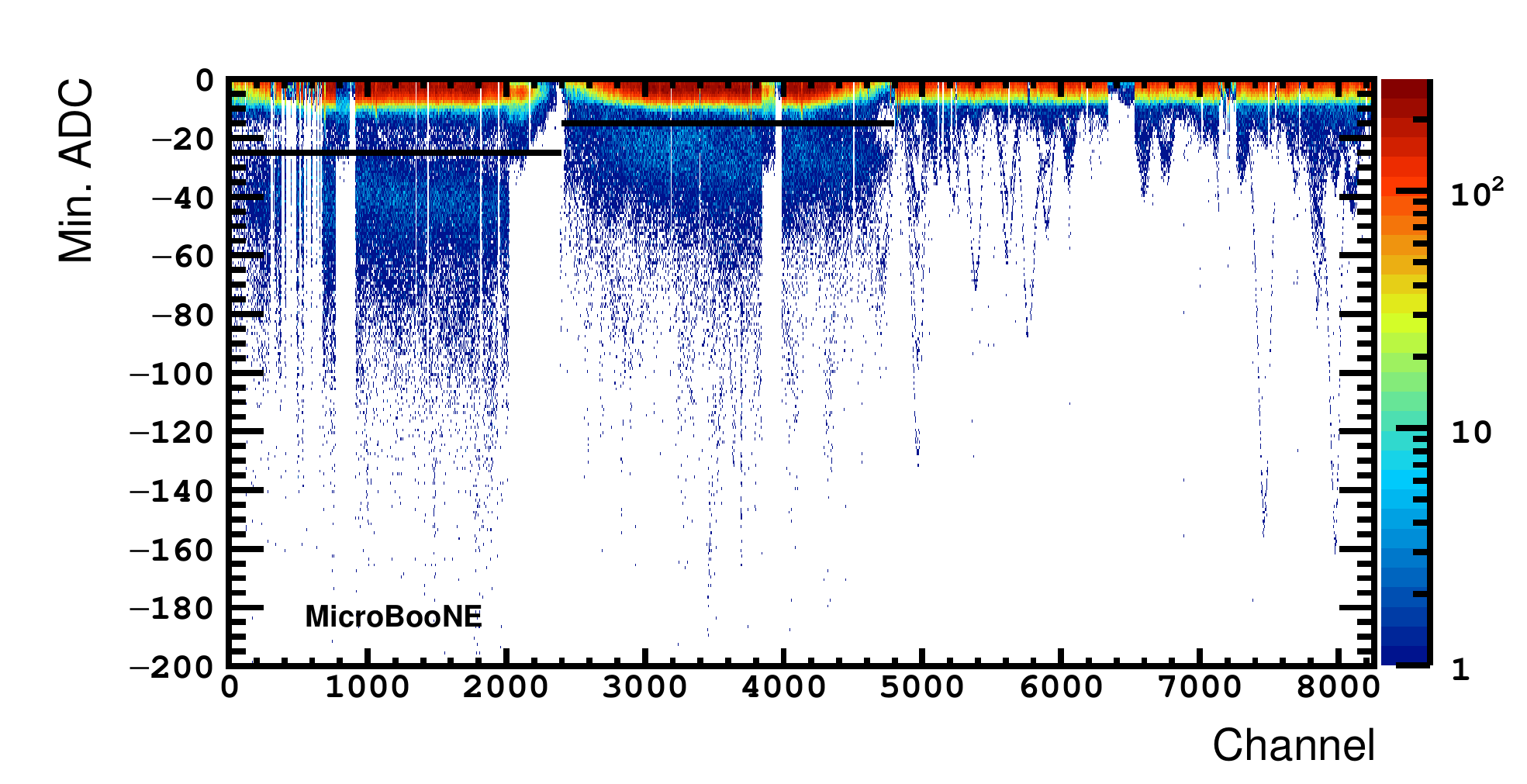}
  \label{fig:ADCSpectraPerPlane_min}
}
\caption[Maximum and minimum ADC values in the regions of interest produced by the LArSoft emulation of the ZS algorithm with a $\pm 5$~ADC count threshold.]{Maximum \subref{fig:ADCSpectraPerPlane_max} and minimum \subref{fig:ADCSpectraPerPlane_min} ADC values in the regions of interest produced by the emulation of the ZS algorithm with a $\pm 5$~ADC count threshold to decimate the SN stream raw data, as a function of channel number. The horizontal black lines mark the location of the chosen thresholds for the three TPC planes: $-25~\rm{ADC}$ counts for plane U (channels $0 - 2399$), $\pm15~\rm{ADC}$ counts for plane V (channels $2400 - 4799$) and $+30~\rm{ADC}$ counts for plane Y (channels $4800 - 8255$).}
\label{fig:ADCSpectraPerPlane}
\end{figure}

Using events from off-beam zero-bias triggers from the trigger stream during previous runs with good electron lifetime, a first pass with a software emulation of the ZS algorithm with a threshold of $\pm 5$~ADC counts (chosen arbitrarily to reduce the raw data while keeping the signals and some noise fluctuations) and dynamic baseline was done.
ADC spectra of the maximum and minimum ADC values found in each zero-suppressed waveform, after offline baseline subtraction using the first sample, 
were produced as shown in figure~\ref{fig:ADCSpectraPerPlane}. For these spectra, the peak closest to the origin is interpreted as noise, while the peak furthest from the origin is interpreted as the signal, dominated by near-MIP cosmic-ray muons. 
The FPGA firmware only admits one threshold value per channel and its polarity. For the U plane, the negative valley was used to set the threshold at $-25~\rm{ADC}$ counts (unipolar negative threshold). For the V plane, the two valleys are found at approximately the same absolute ADC value. Hence, a threshold of $\pm 15$~ADC counts (bipolar threshold) was established. For the Y plane, the valley in the positive ADC distribution was used to set the threshold at $+30~\rm{ADC}$ counts (unipolar positive threshold).

These amplitude thresholds were tested with the dynamic-baseline ZS, with the baseline tolerance parameters set to $2~\rm{ADC}$ counts for the rounded mean and $3\;\rm{ADC}^2$ counts for the truncated variance based on the MicroBooNE noise levels~\cite{Acciarri:2017sde}. In the end, these amplitude thresholds were deprecated in favor of the bandwidth-driven thresholds described in section~\ref{sec:ZeroSuppressionThresholds}, which allow the recording of more data.

\acknowledgments

This document was prepared by the MicroBooNE collaboration using the
resources of the Fermi National Accelerator Laboratory (Fermilab), a
U.S. Department of Energy, Office of Science, HEP User Facility.
Fermilab is managed by Fermi Research Alliance, LLC (FRA), acting
under Contract No. DE-AC02-07CH11359.  MicroBooNE is supported by the
following: the U.S. Department of Energy, Office of Science, Offices
of High Energy Physics and Nuclear Physics; the U.S. National Science
Foundation; the Swiss National Science Foundation; the Science and
Technology Facilities Council (STFC), part of the United Kingdom Research and Innovation;
 and The Royal Society (United Kingdom).  Additional support for the laser
calibration system and cosmic ray tagger was provided by the Albert
Einstein Center for Fundamental Physics, Bern, Switzerland.

\clearpage
\bibliographystyle{JHEP}
\bibliography{references} 

\end{document}